\documentclass[]{aa}
\pdfoutput=1
\usepackage{graphicx}
\usepackage{txfonts}
\begin{document}
%

   \title{ Proper motions for HST observations in three off-axis bulge
     fields\thanks{Based on observations made with the NASA/ESA Hubble Space Telescope, obtained from the data archive at the Space Telescope Institute. STScI is operated by the association of Universities for Research in Astronomy, Inc. under the NASA contract NAS 5-26555. }}

   \author{M.Soto\inst{1,2,3}, H.Zeballos\inst{2,4}, K.Kuijken\inst{2},
     R. M. Rich\inst{4}, A. Kunder \inst{6}
          \and
           T. Astraatmadja\inst{7} }          
   \institute{Space Telescope Science Institute, 
          3700 San Martin Drive, Baltimore, MD 21218, USA.\\
           \email{msoto@stsci.edu}
           \and
              Leiden Observatory, Leiden University
              PO Box 9513, 2300RA Leiden, The Netherlands.\\
              \email{kuijken@strw.leidenuniv.nl}
          \and
            Departamento de F\'{\i}sica, Universidad de La Serena,
            Benavente 980, La Serena, Chile. 
          \and
             Departamento de Astronom\'{\i}a, Universidad de Chile, 
             Casilla 36-D, Santiago, Chile.\\
              \email{zeballos@das.uchile.cl}
          \and
              Division of Astronomy, Department of Physics and Astronomy,
              UCLA, Los Angeles, CA 90095-1562, USA.\\
              \email{rmr@astro.ucla.edu}
          \and
              Leibniz-Institut f\"ur Astrophysik Potsdam (AIP), 
              An der Sternwarte 16, D-14482, Potsdam, Germany\\
              \email{akunder@aip.de}
          \and
          Max Planck Institute for Astronomy, K\"onigstuhl 17, 69117
          Heidelberg, Germany \\
          \email{astraatmadja@mpia.de}
            }

   \date{Received ; accepted}

 
  \abstract
   {} 
   { This is the second in a series of papers that attempt to unveil
     the kinematic structure
     of the Galactic bulge through studying radial velocities and
     proper motions. 
    We report here $\sim15000$ new proper motions for three
    low foreground-extinction off-axis fields of the Galactic bulge. 
%
   }
   {  Proper motions were derived from a combination of
     \emph{Hubble Space Telescope}   Wide Field Planetary Camera 2 (WFPC2) and 
    Advanced Camera for Surveys (ACS) images taken 8 and 9 years apart, 
   and they reach accuracies better than $0.9 \ mas \ yr^{-1}$ for more
   than $\sim 10000$ objects with magnitudes $F814W\leq24$. 
    }
   {The proper motion distributions in these fields are similar to 
    those of Galactic minor axis bulge fields.  
   We observe the 
    rotation of main sequence stars below the turn-off within
      the Galactic bulge, as in the minor axis fields.     
   }
   { Our stellar proper motions measurements show a significant bulge rotation for
     fields as far from the galactic plane as $b\simeq-8^{\circ}$.
  }
   \keywords{Galaxy:bulge --
                Galaxy: kinematics and dynamics --
                Galaxy: stellar content --
                Methods: data analysis
               }
  \titlerunning{Proper motions for HST observations in three off-axis fields}
  \authorrunning{Soto et al.}  
   \maketitle
%

\section{Introduction}
It is still uncertain exactly what mechanisms
led to the formation of the present-day Galactic bulge.
 It is not clear whether the evolution of the bulge was driven by mergers, 
 as the hierarchical galaxy formation scenarios  
 suggest, or secularly by disk instabilities.   
 A clear observational picture of the bulge's current structure is needed to begin to understand
 its formation and evolution.
Because of the high and variable 
 extinction towards the Galactic center, this has been a difficult task and is perhaps the main 
 obstacle to formulating a unified picture. 
  In addition to the high foreground extinction by dust towards the
 Galactic bulge (which is not 
 constant even on small scales), the bulge and the disk are projected on top of each other on the sky. 
 Disentangling them is not straightforward: even in the color-magnitude diagram (CMD), the populations
 overlap (Holtzman et al 1998).
 Moreover, blue stragglers extend brighter than the turn-off and overlap with the main sequence
 region hosting the young population. This complicates the separation of
 populations based on photometry alone, and additional measurements are required to 
 accurately study the different components of the bulge.
 


 Despite these difficulties, progress has been made in understanding the 
 Galactic bulge, especially over the past years.  
 Abundance studies, such as those by Rich (1988), McWilliam \& Rich (1994),  
 Zoccali et al. (2008), and Bensby et al. (2011), have 
 shown that the Galactic bulge has a wide range of metallicities. 
 However, bulge metallicities do typically differ
 from disk and halo populations, showing a comparatively metal-rich population.
 The $\alpha$-elements in the bulge have also been consistently found to be overabundant with respect 
 to halo and disk (Zoccali et al. 2006; Fulbright 2007; Hill et al. 2011).  In particular,
 $\alpha$-elements are related to the formation timescale of the Galactic bulge 
  since they are primarily produced during the explosion of SN Type II (due to short-lived 
 massive stars).  Iron production, on the other hand, is favored by SNe Type Ia explosion, 
 where SN Type Ia typically have a timescale of an order of magnitude
 longer than SN Type II. 
 Therefore, the overabundance of $\alpha$-elements in the bulge suggests a 
 rapid formation scenario. 
 Evidently, most bulge stars were formed before the interstellar medium (ISM) could be
 enriched by SN Type Ia explosions, hence the short inferred formation
 timescale for the bulge ($<$ 1 Gyr)  (e.g. Ballero et al. 2007 and references therein).

 Simulations in the last few years have also complicated our view
 of the formation scenarios of the Galactic bulge. Shen et al. (2010)
 reproduced the stellar kinematics of the Bulge Radial Velocity Assay
 (BRAVA; Rich et al. 2007) without a classical bulge, where the boxy
 bulge previously reported in the literature is the end-on projection
 of the bar structure. Conversely, Saha et al. (2012) finds that a
 non-rotating small classical bulge can evolve secularly through 
 angular-momentum exchange with the Galactic bar.

 At the same time, number counts 
 along the CMD between populous clusters at several latitudes have been used to estimate 
 the foreground disk contamination (\cite{feltzing}) and the age of the
 bulge. The foreground disk population mimics the young 
 bulge population, especially near the turn-off, affecting the age determinations in the bulge. 
 The results for two bulge fields that effectively identified the contamination by foreground
 populations have placed the age of the bulge population as old as $\sim10\ Gyr$.  

\begin{table*}
\begin{center}
\caption{Summary of observations\label{tab:summob}}. 
\small	
\begin{tabular}{ l l l l l l l}
\hline
\hline
Field     &   Epoch & Exp.(s) & Instrument & Filter & Type & $\alpha,\delta \ (J2000)$\\
\hline
Field 4-7 &  1995 July 14 & 1200($\times$2), 1300($\times$2) & WFPC2 & F555W, F814W & Undithered &  18 22 16, -29 19 22  \\
                & 2004 July 11 & 50($\times$2), 347($\times$4) & ACS WFC & F814W & Dithered &  \\
Field 3-8 &  1996 May 1 &  2900($\times$3) & WFPC2  & F555W, F814W & Undithered &  18 24 09, -30 16 12  \\
                                               & 2004 July 12 & 50($\times$2), 348($\times$4) & ACS WFC & F814W & Dithered &  \\
Field 10-8 &  1995 Nov 30 & 1200($\times$4) & WFPC2 & F555W, F814W & Undithered &  18 36 35, -23 57 01  \\  
                                               & 2004 July 14 & 50($\times$2), 347($\times$4) & ACS WFC & F814W & Dithered &  \\
\hline
\end{tabular}
\end{center}
\end{table*}
\normalsize

 Despite its proximity, the Galactic bulge
 has classically suffered from a lack of proper-motion studies. 
 Spaenhauer et al. (1992) were the first to obtain reliable 
 proper motions from a Galactic bulge sample using photometric plates taken more than three decades
 apart. This study was the subject of a subsequent analysis by Zhao et al. (1994), who included
 radial velocities for a small subsample (64 stars). They found  
 a significant vertex deviation (i.e., a triaxility signature) for the metal-rich population in their
 small proper motion-radial velocity  combined
 sample. The same signature of triaxility was observed
 by Soto, Rich \& Kuijken (2007), who combined
 Spaenhauer et al. (1992) proper motions with \cite{sadler} (1996) and
 \cite{terndrup} (1995) radial velocities and metallicities to obtain a sample of $\sim
 300$ K giants with 3-D kinematics and metallicities. 
  More recently, the same triaxility signature related to two
   distinct populations, metal-rich and metal-poor, has been confirmed
   using high-resolution spectra 
   by Babusiaux et al. (2010) and Hill et al. (2011). 

 Microlensing surveys of the Milky Way bulge have also
 contributed to 
 improving our knowledge of the kinematics of the Galactic bulge.
 The OGLE-II experiment has produced 
 $\sim \ 5\times10^6$ proper motions (\cite{sumi}) for 49 bulge fields, covering a range of
 $-11^{\circ} < l < 11^{\circ}$ and $-6^{\circ} < l < 3^{\circ}$, and reaching accuracies 
 of $0.8-3.5\ mas\ yr^{-1}$. Using this survey as a basis, Rattenbury
 et al. (2007a; 2007b) studied the proper motions for a subsample of
 bulge red clump giant stars in 44 fields. Red clump stars were used
 as tracers of bulge density in order to fit a mass density
 distribution for the bulge.  Along the same lines, Vieira et al. (2007)
 delivered proper motions for 21,000 stars in Plaut's window  ($l,b= 0^{\circ},
 -8^{\circ}$) from plates spanning 21 years.
 

 Space-based observations have also played a role in bulge proper-motion 
 observations, boasting the combination of sharper images and reduced blending.
 Anderson \& King (2000; 2003) developed a technique of deriving
 precision astrometry.  Their innovations included an
   effective point spread function (PSF) approach that
   obviates the need to integrate the 
 PSF over pixels when evaluating it for a given image, and an
 empirical distortion correction for WFPC2, later on extended 
 to the ACS Wide-Field Channel (WFC) and ACS High-Resolution Channel
 (HRC). These procedures  were subsequently  
 applied to measuring the component of rotation of 47 Tucanae
 globular cluster (Anderson \& King 2003).  
 Similarly, Kuijken \& Rich (2002; henceforth KR02) used a modification of
 Anderson \& King (2000) approach to be the first to use HST observations 
 to obtain reliable bulge proper motions. KR02's sample targeted
 two low foreground-extinction fields,  Baade's window 
 $(l,b=1.1^{\circ},-3.8^{\circ})$ and Sagittarius I $(l,b=1.3^{\circ},-2.7^{\circ})$.

\begin{figure}
\centering
\includegraphics[width=8cm]{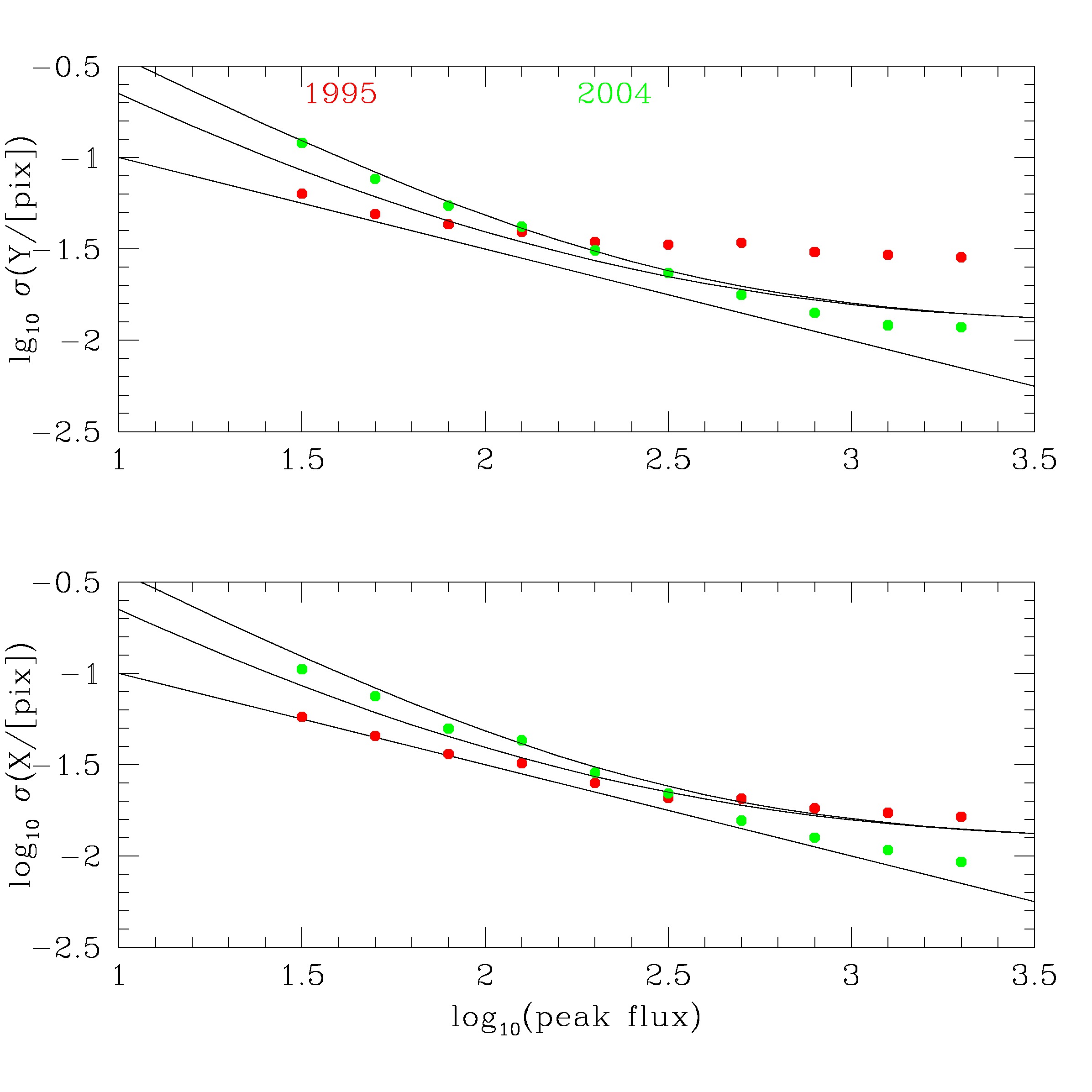}
\caption{The \emph{rms} residuals of the pixels positions in $x$ (bottom)
  and $y$ (top) from the proper
  motion fits in one of our fields, Field 4-7 WF2. 
  The solid lines are the best theoretical fit for an error estimation
  based on photon noise alone (\emph{bottom line}) and for photon noise
  including additional systematic source of errors for each epoch
  (\emph{two top lines}).
 This plot shows
  that the  dominant source of error for the fainter undithered first epoch (\emph{red})
  is photon noise, while for brighter sources a residual systematic error
  appears that we relate to saturation due to the long
  exposures. Second-epoch dithered exposures (\emph{green}) are best
  represented by a systematic error that has been included in the
  estimation of the proper motion error.   
\label{fig:syserror}} 
\end{figure}

  Both fields  
  were used to successfully obtain 15,862 and 20,234 proper motions respectively,
  proving the feasibility of space-based proper motions with
  considerably shorter time baselines
 than those previously employed on ground-based
 bulge proper motions (e.g., Spaenhauer et al. 1992; $\sim30\ yr$). The samples in both fields
 were separated into bulge and disk components based on the mean proper motions.  
 From this it was found that the bulge component clearly resembles an old population, 
 such as those observed in globular 
 clusters, and shows a significant rotation with no covariance in \emph{{l,b}}. 
 More recently, both fields in KR02 have been the subject of new
 proper motion studies.
 Koz\l owski et al. (2006) obtained proper motions for 35 small fields in the
 vicinity of Baade's window. Their calculations, involving the
 combination of ACS/HRC  
 images and 
 archival observations with WFPC2, yielded 15,863 stellar proper motions. 
 The proper motions calculated by  Koz\l owski et al. (2006) show
 consistency with the velocity dispersions found by KR02,  and 
 in addition show a significant negative covariance term in the transverse velocity 
 $C_{lb}=\sigma_{lb}/(\sigma_l \sigma_b)\simeq -0.10$. A negative
 covariance such as this may imply a tilt in 
 the velocity ellipsoid with respect to the Galactic plane. 

 Similarly,
 new results for the  \emph{Sagittarius I} field (\cite{clarkson}) yielded
 more than $180,000$ proper motions with ACS epochs separated by just 2 years. 
 From their initial numbers they
 finally selected 15,323 bulge stars using a similar procedure to the one applied by 
 KR02. The covariance they report was very similar to the one found by Koz\l owski et al. (2006).  
 In addition, Clarkson et al. (2008) produced velocity ellipsoids  in (\emph{l,b}) as a function of 
 distance bins.
 These velocity ellipsoids demonstrate a slight dependence on the
 distance of the objects analyzed.  Similar to the finding in KR02,  
 the median stellar sequence in Clarkson et al. (2008) for this bulge 
 sample was best represented by an 11 Gyr
 old isochrone.

 We report here new proper motions results for stars in
 three low foreground-extinction windows of the Galactic bulge located
 in the first Galactic quadrant.
 These new fields differ from other HST bulge proper motion
  observations, such as the minor-axis fields presented in Clarkson et al. (2008) and KR02,
  since they sample the near end of the bar in the Galactic bulge. In addition, the derived
  proper motions have been obtained from cross-detector observations; 
 suitable first-epoch WFPC2 exposures were obtained from the HST
 archive, and complemented with more recent ACS WFC
 exposures. 
 These second-epoch observations are part of a larger
   program (GO-9816) that includes observations in both ends of the
  Galactic bar, as well as minor axis fields (see \cite{soto12} for
  more  details). 
 The resulting time baselines for the proper motions 
 in this work are eight to nine years.   

 This paper is organized as follows. In \S 2  we describe our observations. 
 A detailed account of the procedures involved in the measurements 
 of the proper motion can be found in \S 3. Section 4 presents the 
 analysis performed on our sample of proper motions 
 and the implications of the 
 results. In addition, we compare our results with those of similar surveys in the 
 Galactic bulge. Finally, our conclusions are summarized in \S 5.

\begin{figure*}
\centering
\includegraphics[width=6.5cm, angle=0]{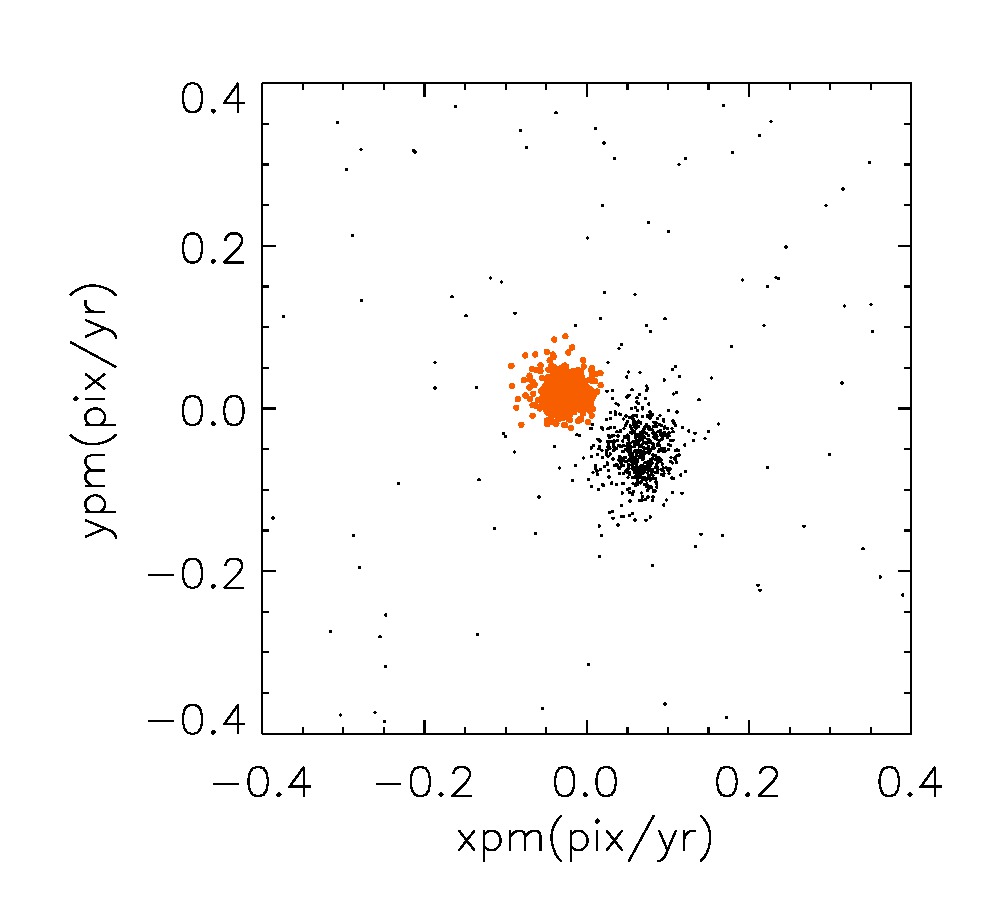}\\
\includegraphics[width=4.2cm, angle=0]{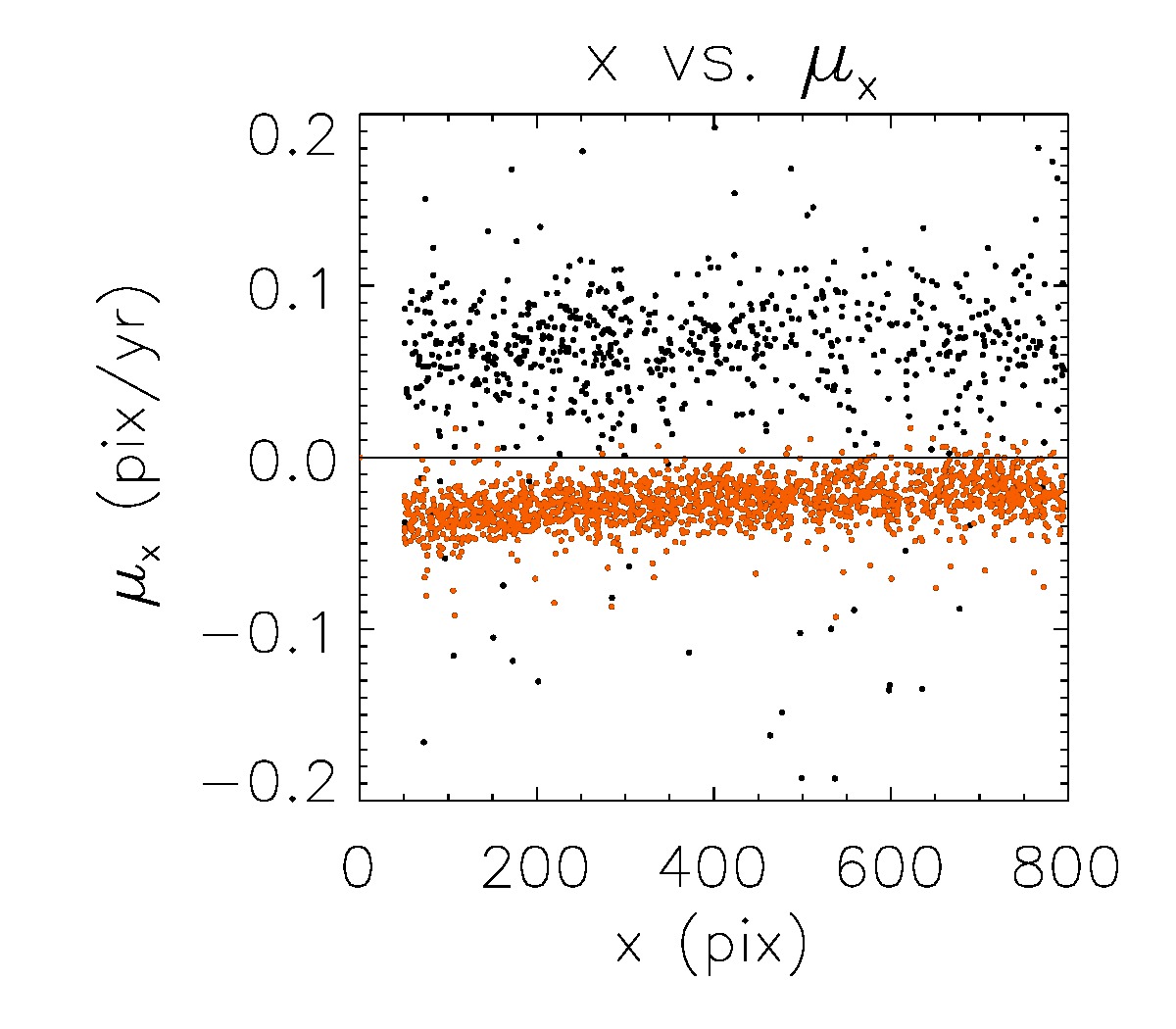}
\includegraphics[width=4.2cm, angle=0]{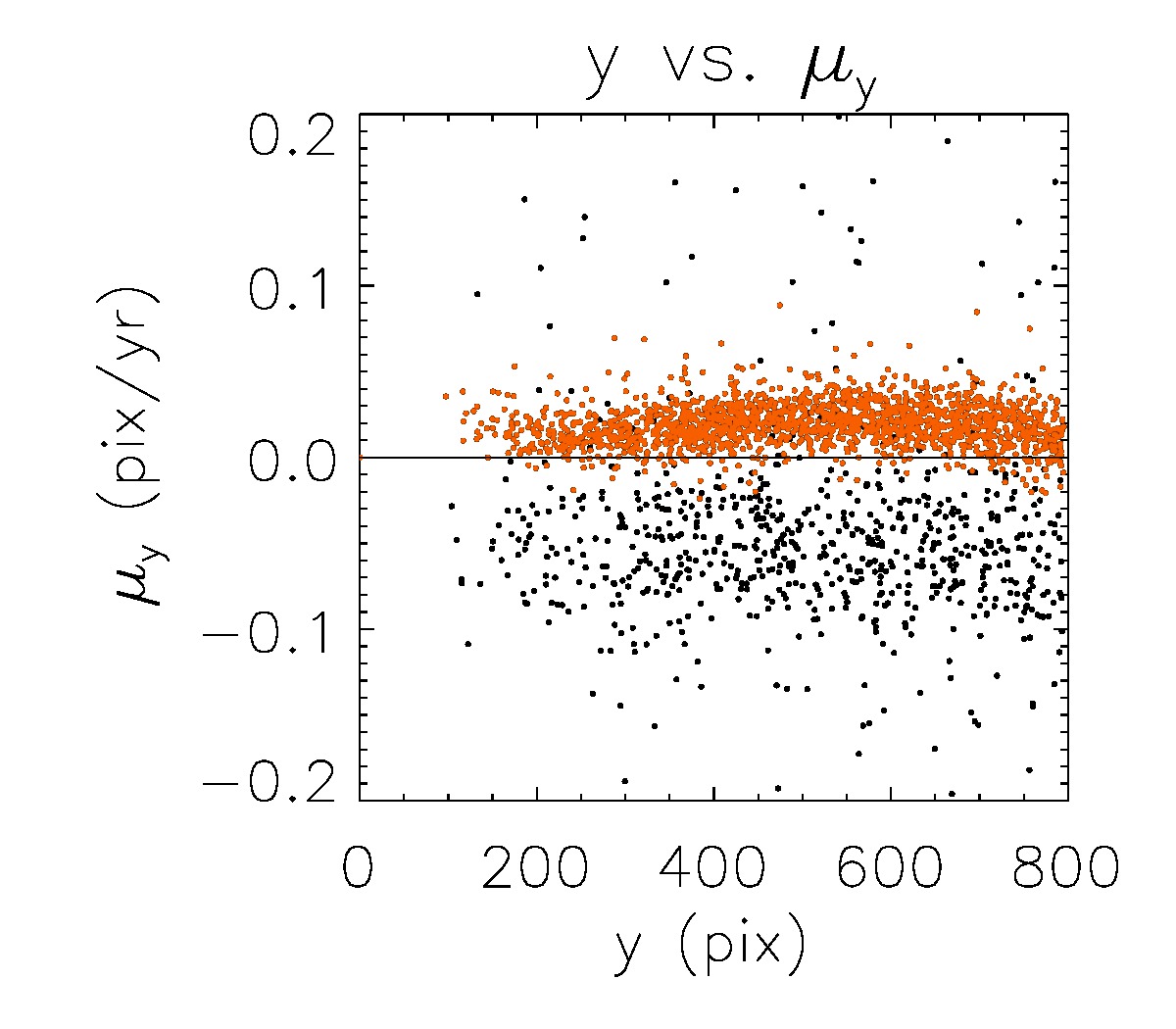}
\includegraphics[width=4.2cm, angle=0]{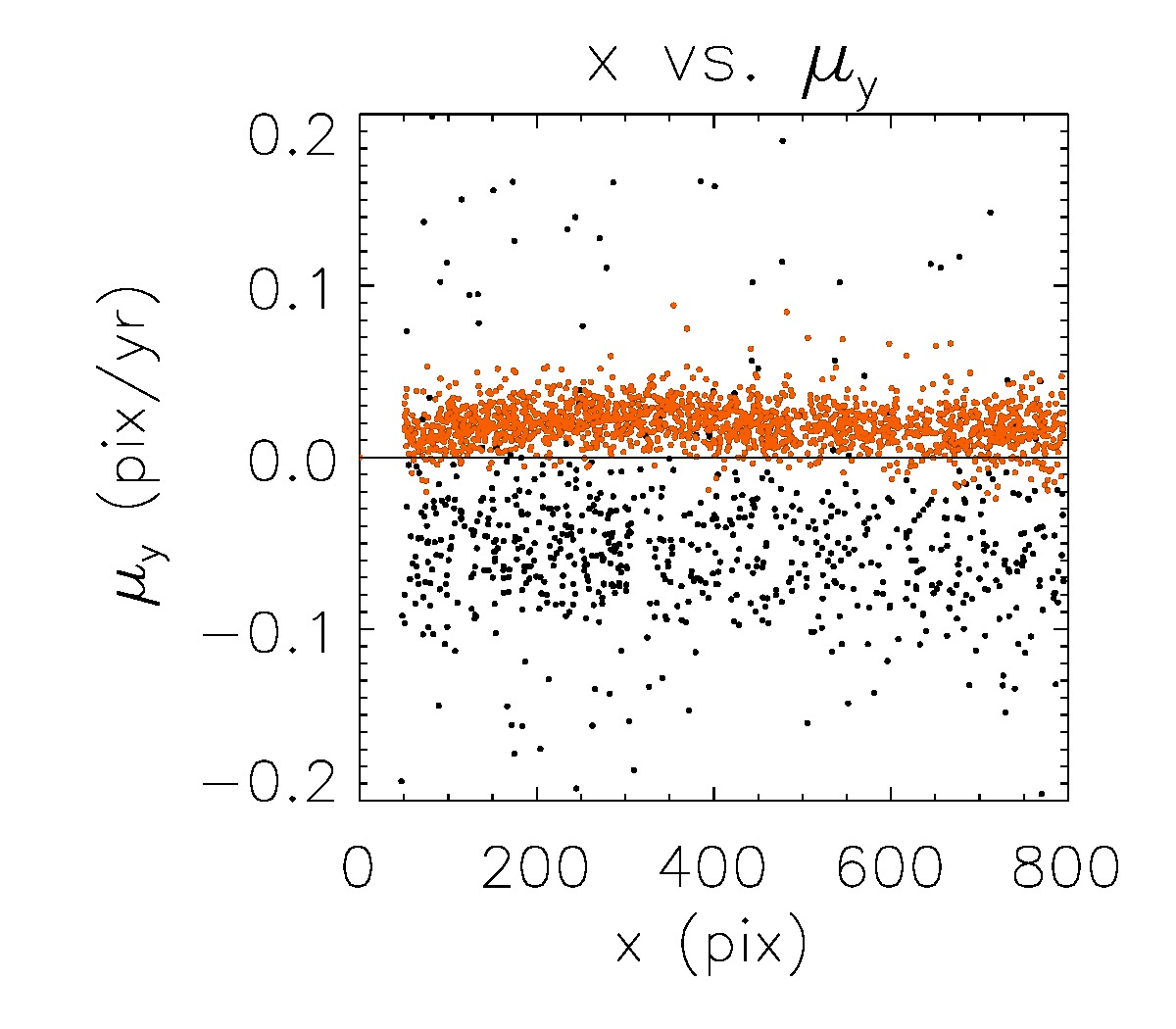}
\includegraphics[width=4.2cm, angle=0]{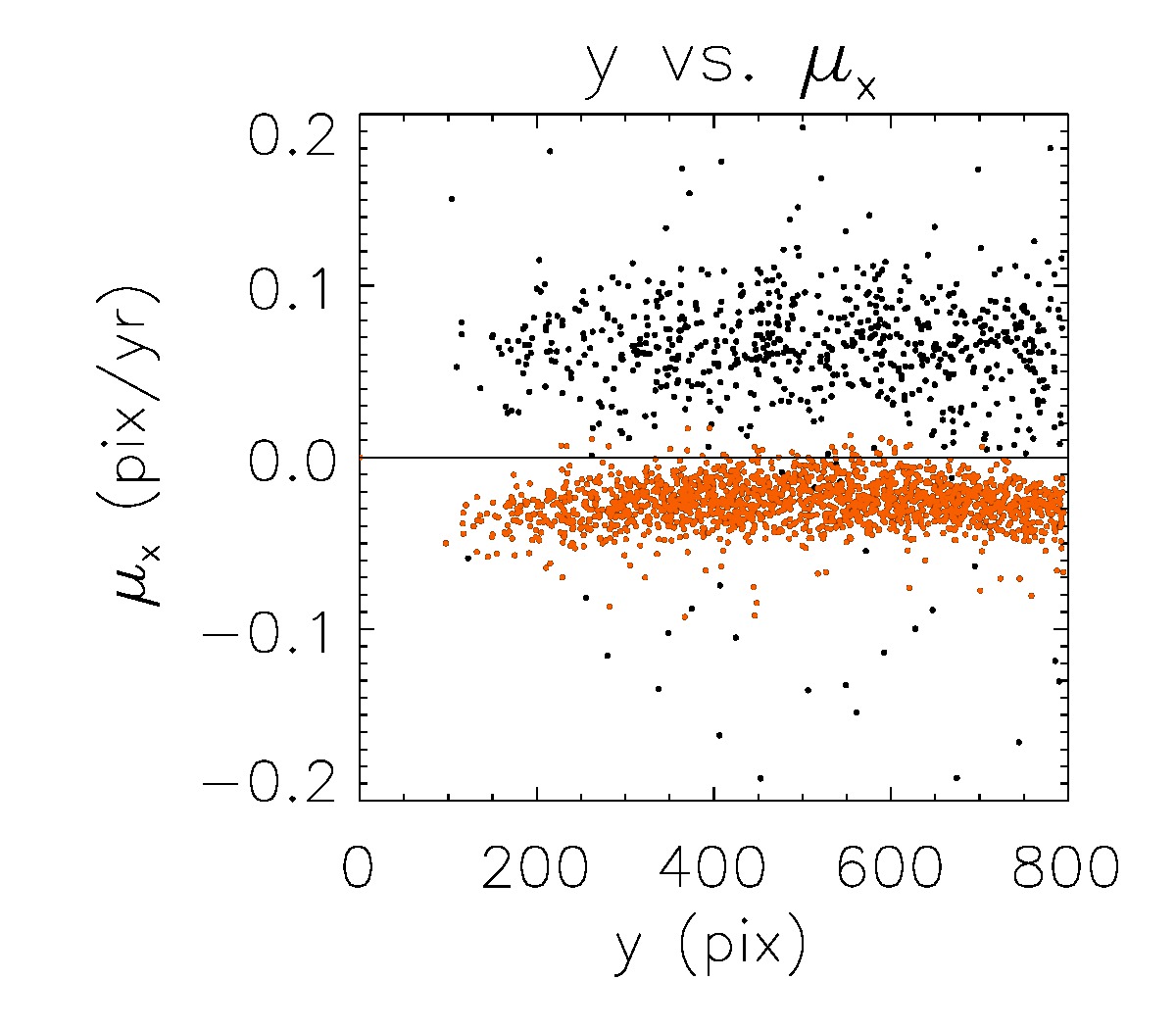}\\
\includegraphics[width=4.2cm, angle=0]{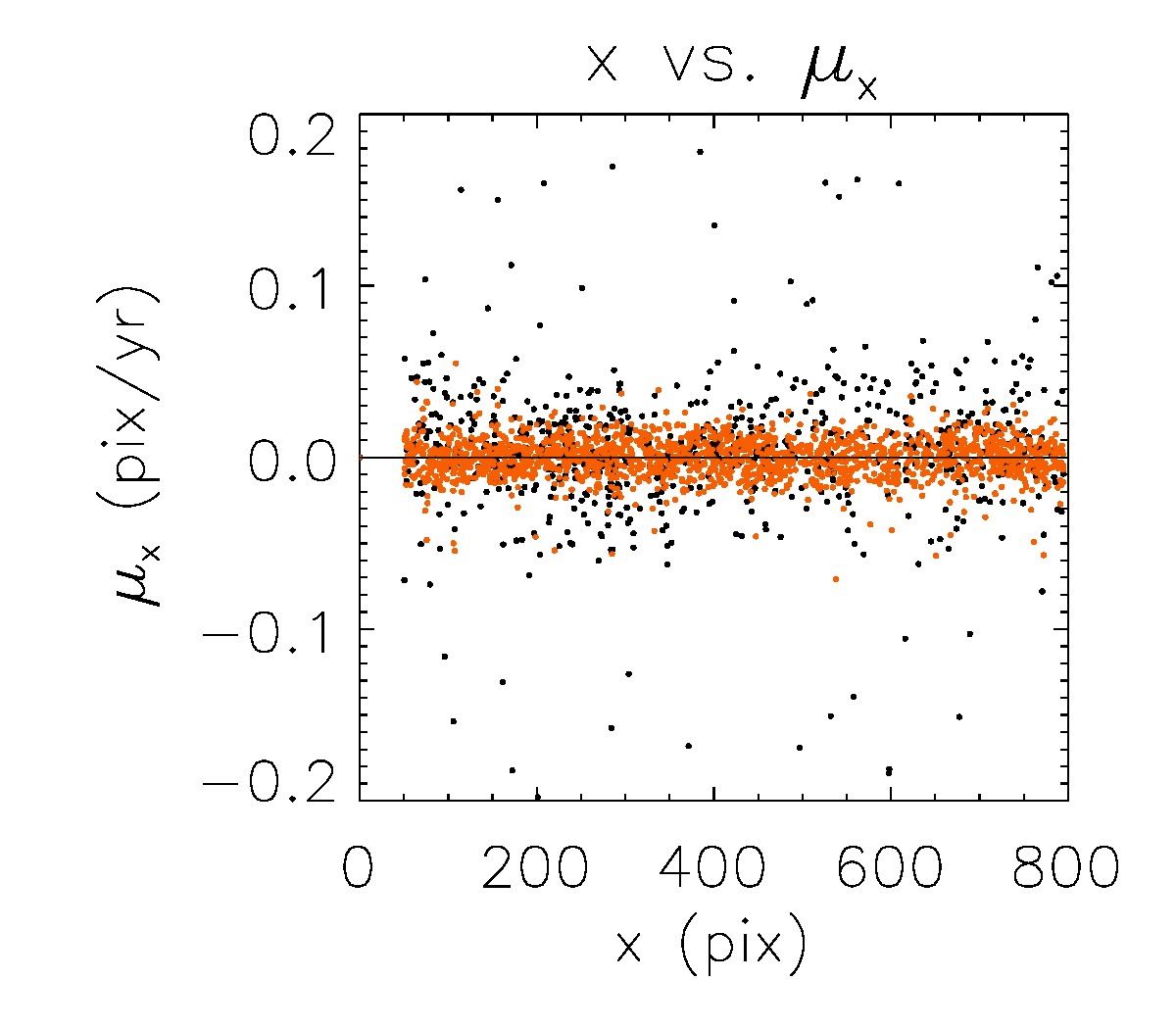}
\includegraphics[width=4.2cm, angle=0]{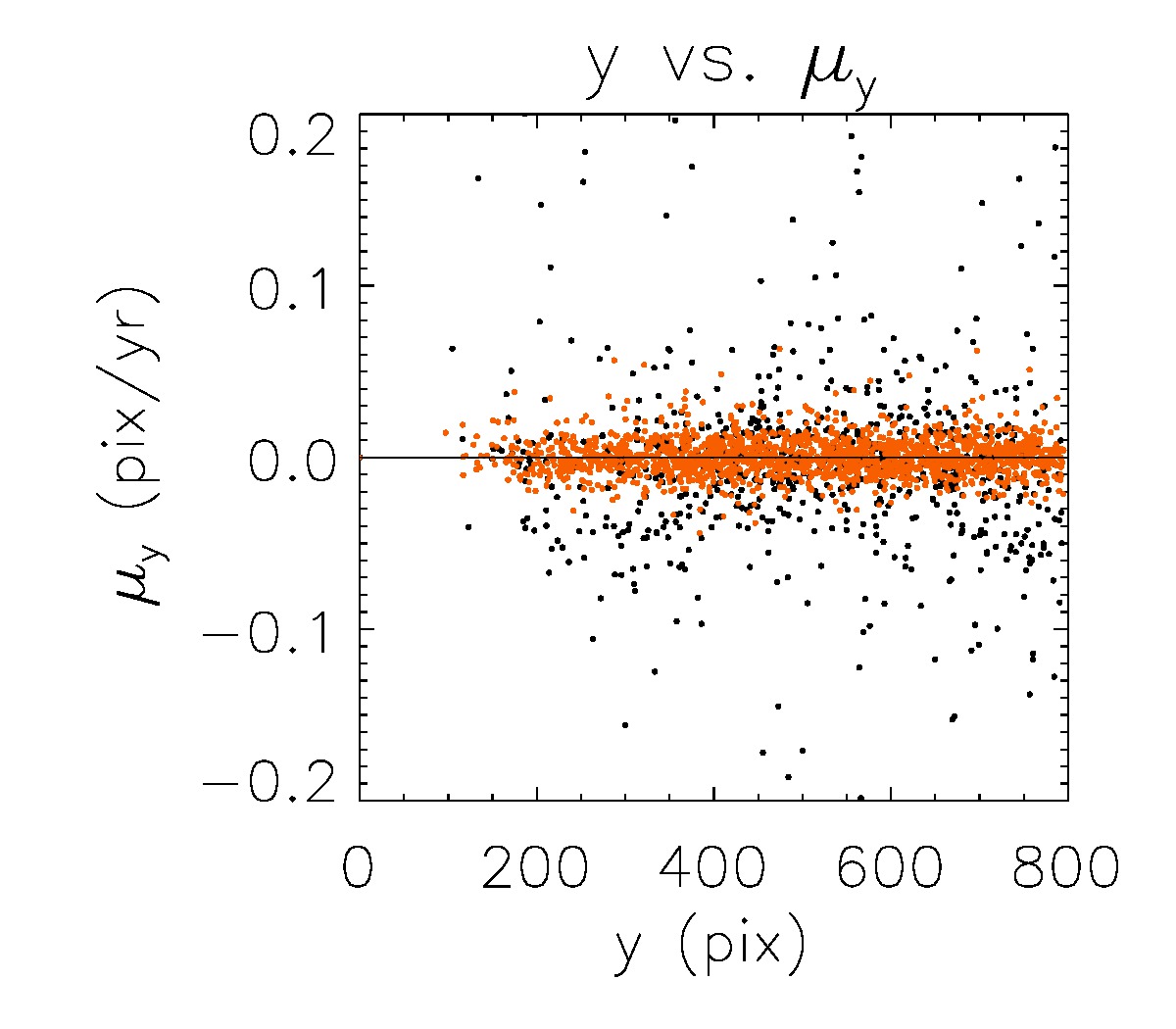}
\includegraphics[width=4.2cm, angle=0]{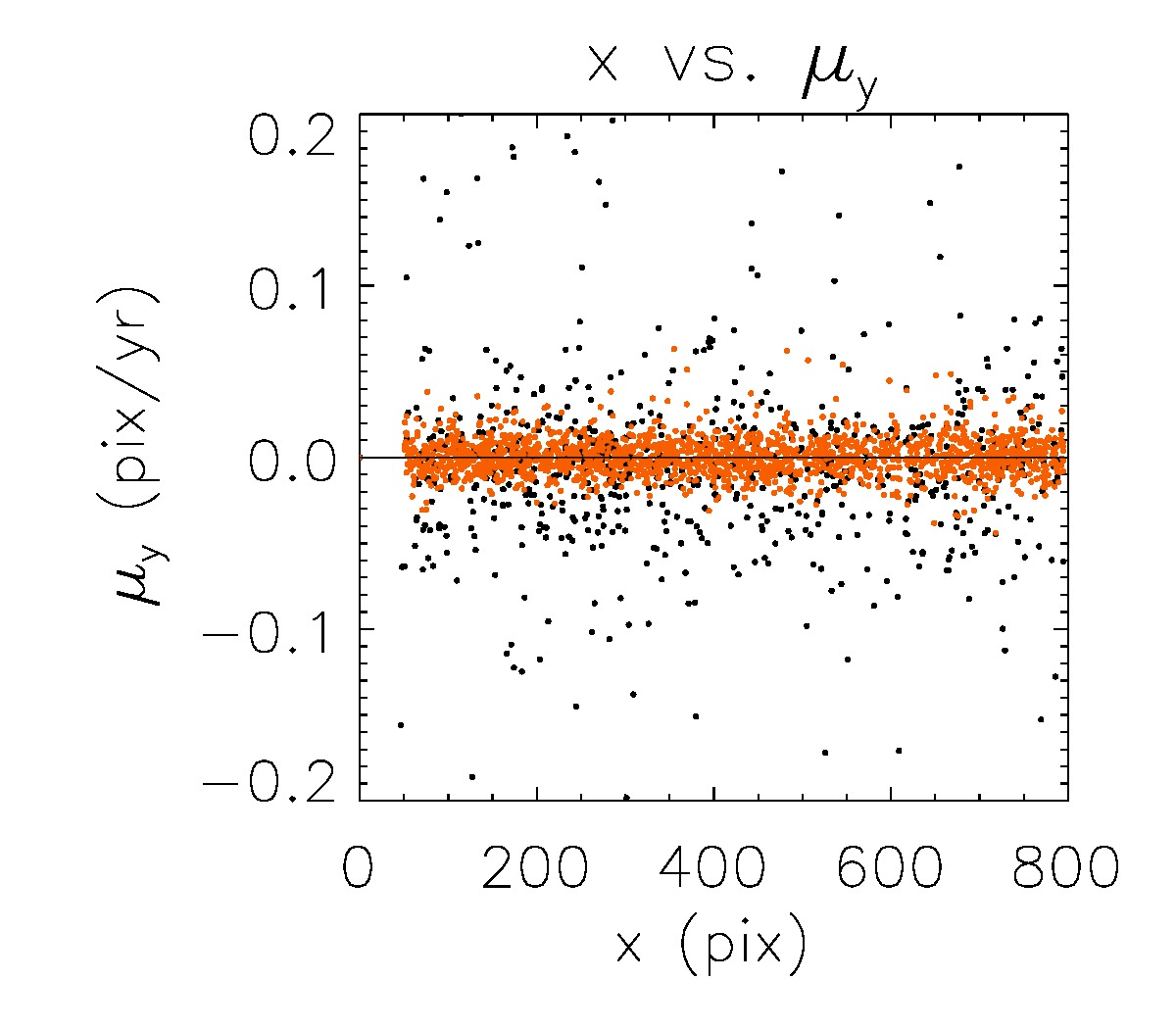}
\includegraphics[width=4.2cm, angle=0]{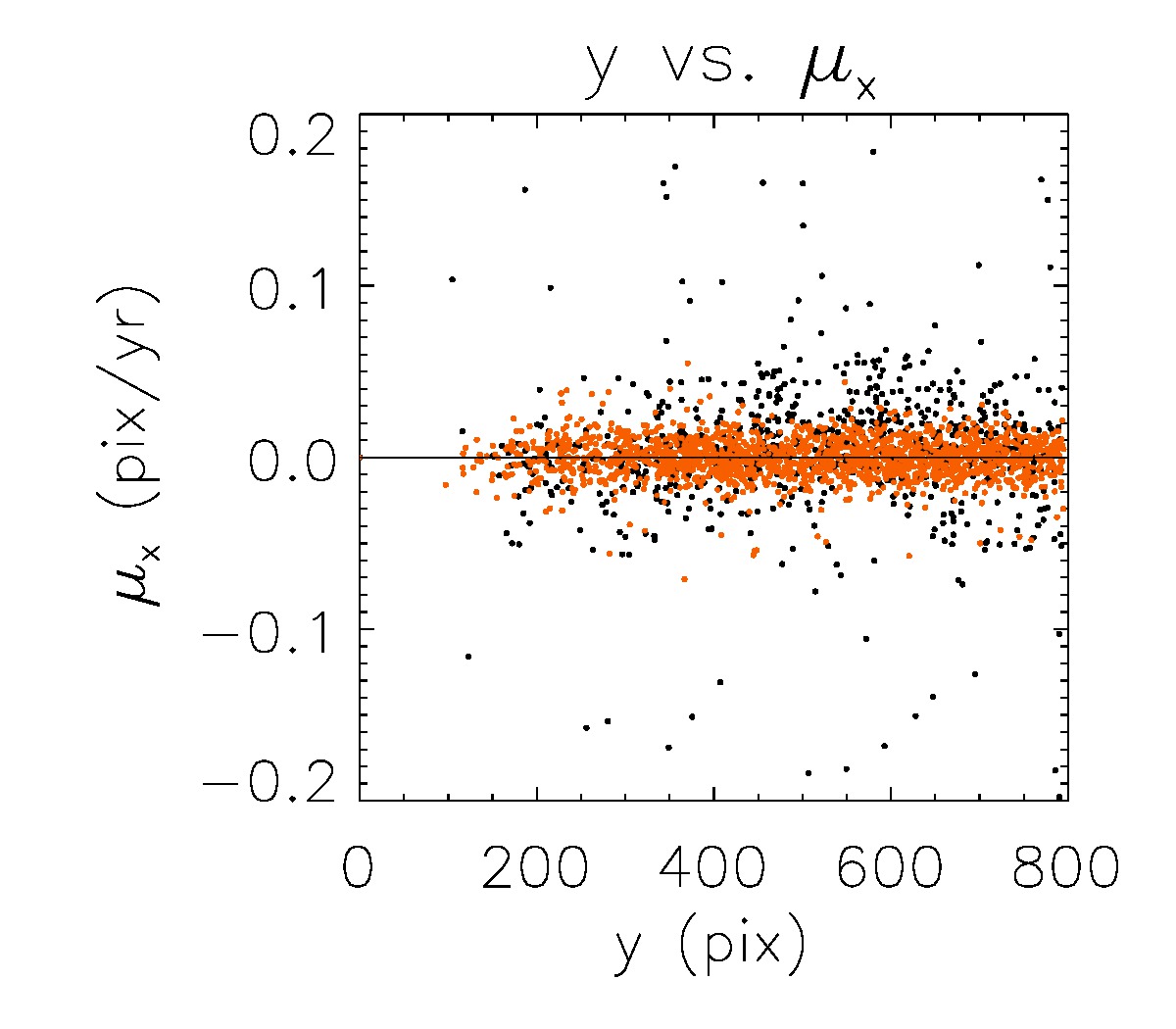}
\caption{Kinematic selection of cluster stars in Field 10-8
    in one of the cameras of WFPC2, Wide Field 4 (WF4). 
  The selection was performed for each chip separately using a hard cut-off; 
 cluster stars have a small velocity dispersion, which therefore biases the relative zeropoint 
 of the proper motions of bulge stars. 
 Thus, the cluster stars must be excluded from the reference frame used to calculate 
 bulge proper motions. 
 \emph{Top row:} proper motions in $pixels \ yr^{-1}$ before the
 refinement of the stars used as the basis for the inter-epoch
  transformations 
 for WF4 chip in Field 10-8; where black dots are bulge stars, and orange dots are
 those kinematically classified as cluster stars. 
 \emph{Second row:} proper motions in pixels $\mu_x$ and $\mu_y$ 
 as a function of the $x-y$ coordinates before the correction. 
 \emph{Bottom row:} same as second row, but after applying the correction. 
\label{fig:pm2coor_wf4}} 
\end{figure*}

\section{Observations}


 Our observations target three fields at positive Galactic longitudes; 
 Field 4-7 [$(l,b)=(3.58^{\circ},-7.17^{\circ})$], 
 Field 3-8 [$(l,b)=(2.91^{\circ},-7.96^{\circ})$], and 
 Field 10-8 [$(l,b)=(9.86^{\circ},-7.61^{\circ})$]. 
 First-epoch observations were obtained from the Hubble Space
 Telescope (HST) data archive, which is now over a decade and a half old. 
 In addition to the three fields analyzed here, 
 this wealth of images also provided the basis for the three low
 foreground-extinction fields close to the Galactic minor axis;
 these results are reported in KR02 and Kuijken (2004; henceforth K04). 
  Two aspects of the first-epoch data are not ideal: (i) the exposures are all long ($> 1200 s$), so 
  the positions of the bright stars are not measurable; (ii) the exposures are not dithered, 
  which limits our ability to build a PSF model.
  To mitigate the 
  effect of the undersampling of the PSF, we chose the F814W
  filter that has the widest, hence the least, undersampled PSF.

The observations for our three fields have combined WFPC2 and ACS WFC
for the first and
 second epochs, respectively. ACS WFC observations were preferred for
 the second-epoch,
 in spite of the cross-instrument systematics, owing to their
 extended field and the resolution which an
 eventual third ACS epoch could take advantage of.
 The ACS WFC second-epoch observations were acquired during 2004 
 July 12 and 14. 
 Two short exposures with $50\ s$  of integration time 
 and four longer exposures of $\sim350 \ s$ were taken for each field.  
 These observations were dithered using a line pattern for the
   $50\ s$ exposures, while a box pattern was chosen for the longer
   $347\ s$ exposures, and spacings in each case were set to
   ($0.1825$) arcsec and  ($0.265,0.187$) arcsec, respectively. 
 These ACS second-epoch observations are
 not completely aligned with the WFPC2 first-epoch. 
 Inclination angles of  $\sim14^{\circ}$ for fields Field 4-7 and
 Field 3-8, and $\sim17^{\circ}$ for Field 10-8 were measured. 
 Table \ref{tab:summob} summarizes our observation in both epochs for our three fields. 

 It has been demonstrated that for a wide range of realistic PSF profiles, the 1$\sigma$ uncertainty
 in the centroid of a stellar images of full width at half maximum, $FWHM$, and signal-to-noise 
 ratio, $S/N$, is given by $\simeq 0.7\times FWHM/(S/N)$ (KR02). 
 This theoretical limit gives us a reference for what can be expected 
 in our proper motions,
 for first-epoch observations were performed in WFPC2 ($FWHM \simeq 0.12"$ in F814W) a star
 detected at 15 $\sigma$ can be centered with a precision of $5.6 \ mas$. Therefore, a time 
 baseline of 8 yr (in our fields we have 8-9 yr of baseline) yields an accuracy of 
 $0.53\ mas\ yr^{-1}$, which corresponds to $\sim 20 \ km \ s^{-1}$ at
 the distance of the bulge ($\sim8$ kpc).  Since our second-epoch observations are dithered  
 and are of a higher resolution, they reach higher accuracies than those 
 of the first  epoch. However, the latter cannot improve our proper motion
 accuracy beyond the limits of the first epoch, and the spatial variations 
 in the PSF within the ACS WFC image (\cite{anderson06}) have not been considered in 
 our analysis. These PSF spatial variations in ACS, if ignored, can produce $10 \%$ of error in 
 core photometry, which is reduced when a larger aperture is taken. In the case of astrometric
 measurements, the effects are at the 0.01 pixel level ($\simeq 0.001"$ for ACS), which 
 is negligible considering the error introduced by the first epoch.      
 New measurements including only ACS measurements will include this effect 
 in our astrometric measurements in the future. 
 Similarly, another possible source of uncertainty in our proper motion
   solution can be the Charge Transfer Efficiency (CTE) problem in WFPC2
   (Dolphin 2000) and ACS (Kozhurina-Platais et al. 2007).  The CTE
   effect has been carefully characterized for both cameras, and it is
   known to affect the astrometric precision, as well as the
   photometric accuracy. However, this effect should
  not be significant for images taken during 1994-1995 (WFPC2) and
  2004 (ACS), where the instruments had less than $\sim$ 1  and $\sim$
  2 $yr$ in orbit , respectively.  This has also been demonstrated
  for the proper motions of KR02 and Clarkson et al. (2008)
  for other bulge fields.

\begin{figure}[!b]
\centering
\includegraphics[width=6.cm, angle=0]{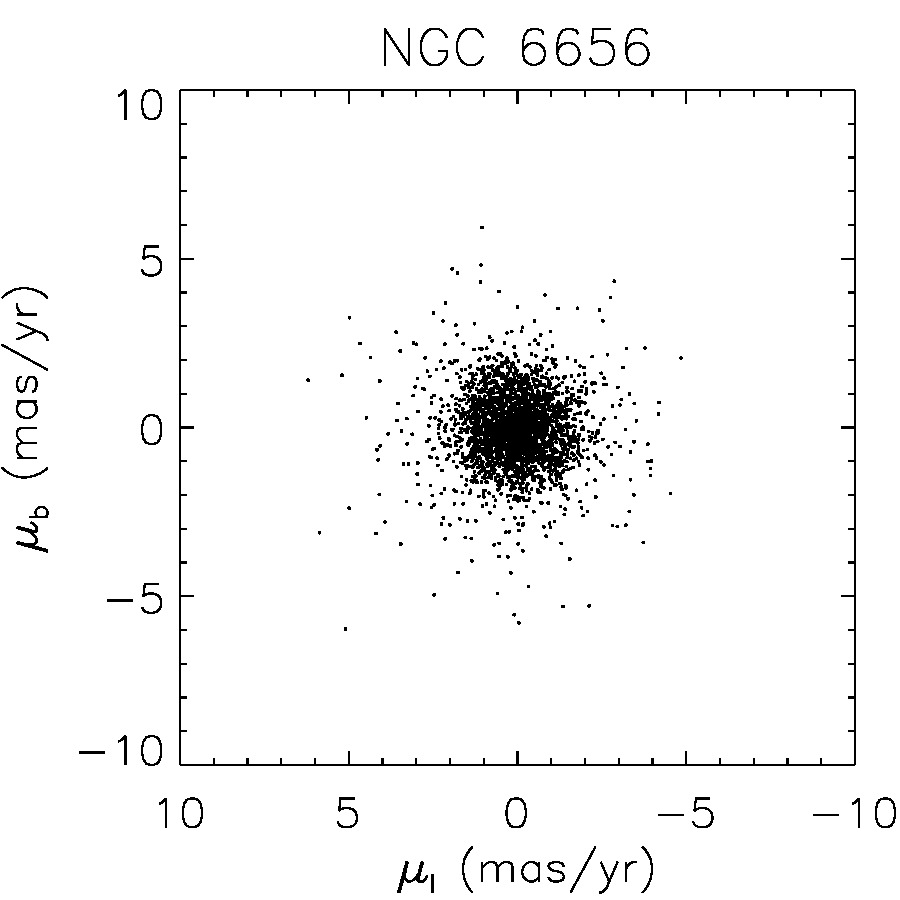}
\caption{ Proper motion distribution for stars belonging to the globular cluster
 NGC 6656 observed in Field 10-8. 
\label{fig:ngc6656pm}} 
\end{figure}

\section{Proper motion measurements}
 Anderson \& King (2000) describe a procedure for accurate 
 astrometric measurements using HST WFPC2, and they subsequently
 updated this technique for ACS WFC (Anderson \& King 2006). Originally these techniques 
 were designed to work with both epochs dithered, where a common PSF model is used     
 for all the dithered exposures. This PSF must be consistent with all the exposures and
 is used to obtain the centroids of each star. Our data, on the other
 hand, with its undithered first-epoch 
 observations, have required some modifications to the original Anderson \& King (2000) 
 recipe. To improve the centroid determination, we determine a PSF 
 for each exposure starting from an analytical model. This model consists of 
 a Gaussian PSF multiplied by a polynomial, where the number of the components for 
 the polynomial accounts for the wings of the PSF and other secondary
 features: three polynomial components in a three-pixel
   fitting radius for WFPC2/WF exposures, and 12 polynomial components
   in a five-pixel 
 fitting radius for WFPC2/PC or ACS/WF exposures. 
 For each exposure, therefore, the PSF starts from this analytical model and is fitted to the 100 
 brightest, unsaturated stars in the image. Our process then iterates using a higher order 
 polynomial, with an robust outlier rejection, which refines the selection until it converges. 
 Typically, a minimum of ten iterations have been run to fit
   the PSF model. 

\begin{center}
\begin{table*}
\centerline{\begin{minipage}[t]{0.7\textwidth}
\caption{Proper motion dispersions in NGC 6656}
\label{tab:velnumdisp2}
\centering
\renewcommand{\footnoterule}{}  
\begin{tabular}{c  l l l c c}
\hline \hline
 & \multicolumn{5}{c}{NGC 6656} \\
\cline{2-6} 
Field    & N & \multicolumn{1}{c}{$\sigma_l$} &
\multicolumn{1}{c}{$\sigma_b$} & \multicolumn{1}{c}{$\sigma_{lb}^2$} &
\multicolumn{1}{c}{$r_{lb}\ $\footnote{Note: $r_{lb}$ is the Pearson correlation coefficient}} \\
         &   & $(mas \ yr^{-1})$ & $(mas \ yr^{-1})$ & $(mas^2 \ yr^{-2})$ &   \\
\hline
PC  & 303  & 0.52$\pm$0.02 & 0.56$\pm$0.04 & -0.01$\pm$0.10 &  -0.01$\pm$0.04 \\
WF2 & 1299  & 0.97$\pm$0.04 & 0.96$\pm$0.02 & 0.11$\pm$0.14 &  0.01$\pm$0.04 \\
WF3 & 2168 & 1.07$\pm$0.02 & 1.05$\pm$0.01 & 0.12$\pm$0.14 &  0.01$\pm$0.02 \\
WF4 & 1678  & 1.08$\pm$0.02 & 1.11$\pm$0.03 &-0.26$\pm$0.06 & -0.06$\pm$0.04 \\
All & 5448 & 1.03$\pm$0.01 & 1.03$\pm$0.01 & -0.11$\pm$0.08 &  -0.01$\pm$0.02 \\
\hline
\end{tabular}
\end{minipage}}
\end{table*}
\end{center}
\normalsize

\begin{figure*}
\centering
\includegraphics[width=15cm]{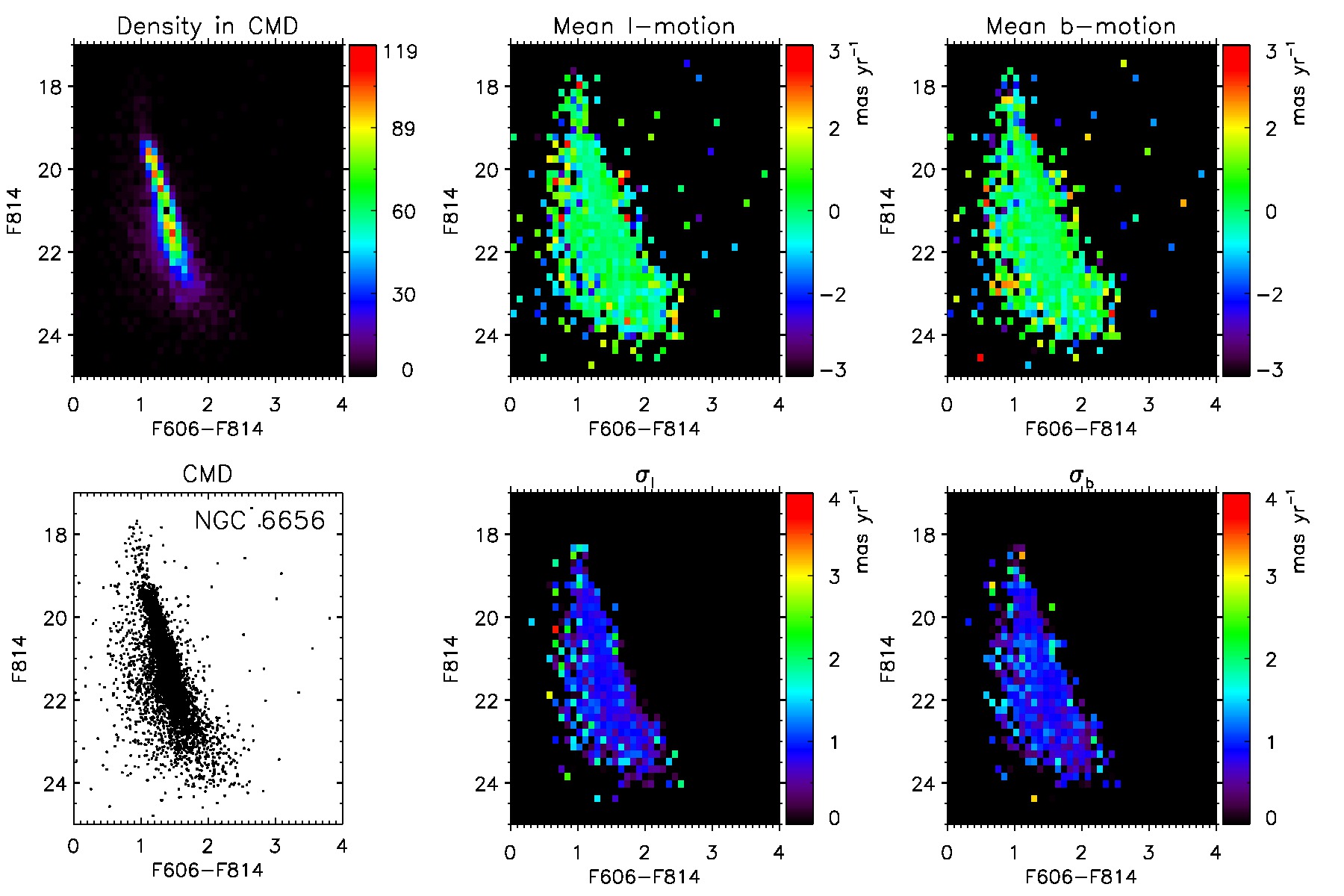}
\caption{Binned CMDs for NGC 6656 stars in Field 10-8
  $(l,b)=(2.91^{\circ},-7.96^{\circ})$ with $n_{fit} \geq 6$. The stars have been selected kinematically using the small kinematic dispersion of the cluster
 proper motions.    
The parameter n$_\mathrm{fit}$ corresponds 
 to the minimum number of exposures fitted by the proper motion
 solution of each star. 
 Some plots have been color-coded using the proper motions and derived dispersions.
 \emph{Top row, left to right}: Stellar density, mean longitudinal proper motion, mean 
 latitudinal proper motions. \emph{Bottom, left to right}:  Unbinned
 CMD,
 longitudinal dispersion
 in proper motions, latitudinal dispersion in proper motions.  
\label{fig:f183633_cmd_b}} 
\end{figure*}

 Once the PSF has been calculated for each exposure, the proper 
 motions can be determined. These are
 obtained starting from a master image that has been generated by
 combining all the flat-fielded \emph{*\_flt} 
 WFPC2, first-epoch (undithered) exposures in the F814W filter.  The
 combined master image is then run through DAOFIND 
 to detect stars with fluxes over $15\sigma$, which produces a master list of stars.
 This master list is used to detect the same stars in all the exposures
 in both epochs; 
 for each exposure, the master list coordinates are used as an initial
 guess to determine the star position.
 In the case of WFPC2 first-epoch exposures, a simple linear transformation
 is sufficient to transform the master-list coordinates to each exposure (e.g., KR02). 
 On the other hand, for our second-epoch ACS WFC images, we have
 used a third-order polynomial transformation, constructed with the
 IRAF task GEOMAP, to convert the
   master-list WFPC2 positions to the ACS/WFC second-epoch exposures.
 This procedure does not affect the final proper motion result as
 long as the transformed master-list coordinates for the second-epoch
 allow a unique identification of the same star in all the second-epoch exposures.  
 For each star, a PSF fit is then computed for the initial positions
  derived from the converted master list coordinates,  to give a best fitting 
 position 
 in each exposure of both epochs.

\begin{table*}
\begin{minipage}[t]{\textwidth}
\caption{Proper motion dispersions}
\label{tab:velnumdisp}
\centering
\renewcommand{\footnoterule}{}  
\begin{tabular}{l l l c c c c}
\hline \hline
Field  & Chip & N & $\sigma_l$ & $\sigma_b$ & $\sigma_{lb}^2$ & $r_{lb}$\\
       &      &   & ($mas\ yr^{-1}$) & ($mas\ yr^{-1}$)  & ($mas^{2}\ yr^{-2}$) &  \\
\hline
Field 4-7  & PC  & 264  & 3.03$\pm$0.21 & 2.98$\pm$0.19  & 0.42$\pm$0.95 & 0.02$\pm$0.11 \\
(nfit$\geq$6)  & WF2 & 2374 & 2.99$\pm$0.06 & 2.58$\pm$0.04  & 0.49$\pm$0.19 & 0.03$\pm$0.03 \\
           & WF3 & 2387 & 3.10$\pm$0.09 & 2.81$\pm$0.05  & 0.93$\pm$0.16 & 0.10$\pm$0.03 \\
           & WF4 & 2360 & 3.01$\pm$0.04 & 2.79$\pm$0.03  & 0.29$\pm$0.46 & 0.01$\pm$0.03 \\
           & All & 7385 & 3.04$\pm$0.02 & 2.74$\pm$0.02  & 0.63$\pm$0.04 & 0.05$\pm$0.01 \\
\hline
Field 3-8  & PC   & 83   & 3.33$\pm$0.45 & 3.53$\pm$0.32 & 0.39$\pm$1.25 & 0.01$\pm$0.18 \\
(nfit$\geq$5)   & WF2  & 1802 & 3.13$\pm$0.09 & 2.98$\pm$0.08 & 0.76$\pm$0.38 & 0.06$\pm$0.04 \\
           & WF3  & 1806 & 3.17$\pm$0.11 & 2.99$\pm$0.07 & 0.71$\pm$0.33 & 0.05$\pm$0.04 \\
           & WF4  & 1231 & 3.08$\pm$0.09 & 2.87$\pm$0.08 &-0.11$\pm$0.57 &-0.01$\pm$0.03 \\
           & All  & 4922 & 3.18$\pm$0.03 & 2.97$\pm$0.04 & 0.69$\pm$0.12 & 0.05$\pm$0.02 \\
\hline
Field 10-8      & PC   & 119   & 2.48$\pm$0.17 & 3.10$\pm$0.49 & 1.31$\pm$0.55 & 0.23$\pm$0.13 \\
(nfit$\geq$6)  & WF2  & 887  & 2.88$\pm$0.11 & 2.71$\pm$0.10 &0.59$\pm$0.51 &0.04$\pm$0.05 \\
                      & WF3  & 855  & 3.11$\pm$0.14 & 2.84$\pm$0.12 & 0.41$\pm$0.75 & 0.02$\pm$0.05 \\
                      & WF4  & 569  & 3.71$\pm$0.18 & 3.52$\pm$0.16 & 1.39$\pm$0.33 & 0.15$\pm$0.06 \\
                      & All  & 2430 & 3.16$\pm$0.06 & 2.98$\pm$0.05 & 0.85$\pm$0.32 & 0.08$\pm$0.02 \\
\hline
\end{tabular}

\end{minipage}
\end{table*}

\begin{figure*}
\centering
\includegraphics[width=6.5cm, angle=0]{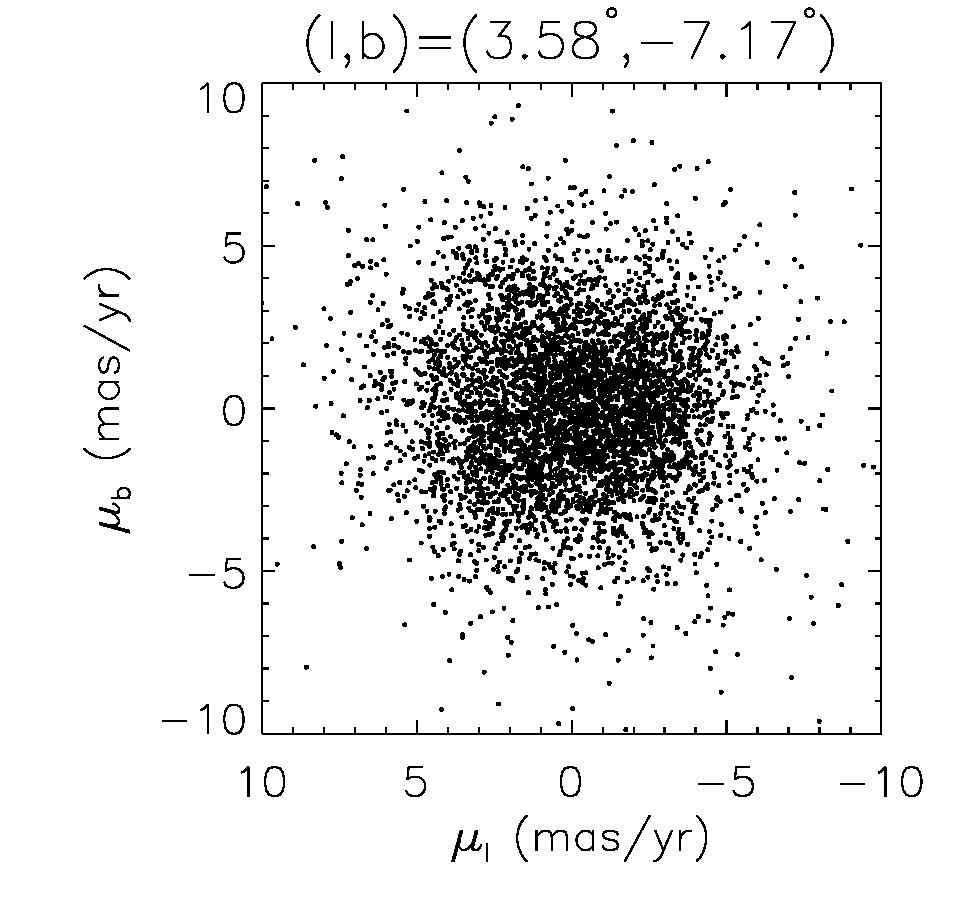}\\
\includegraphics[width=6.5cm, angle=0]{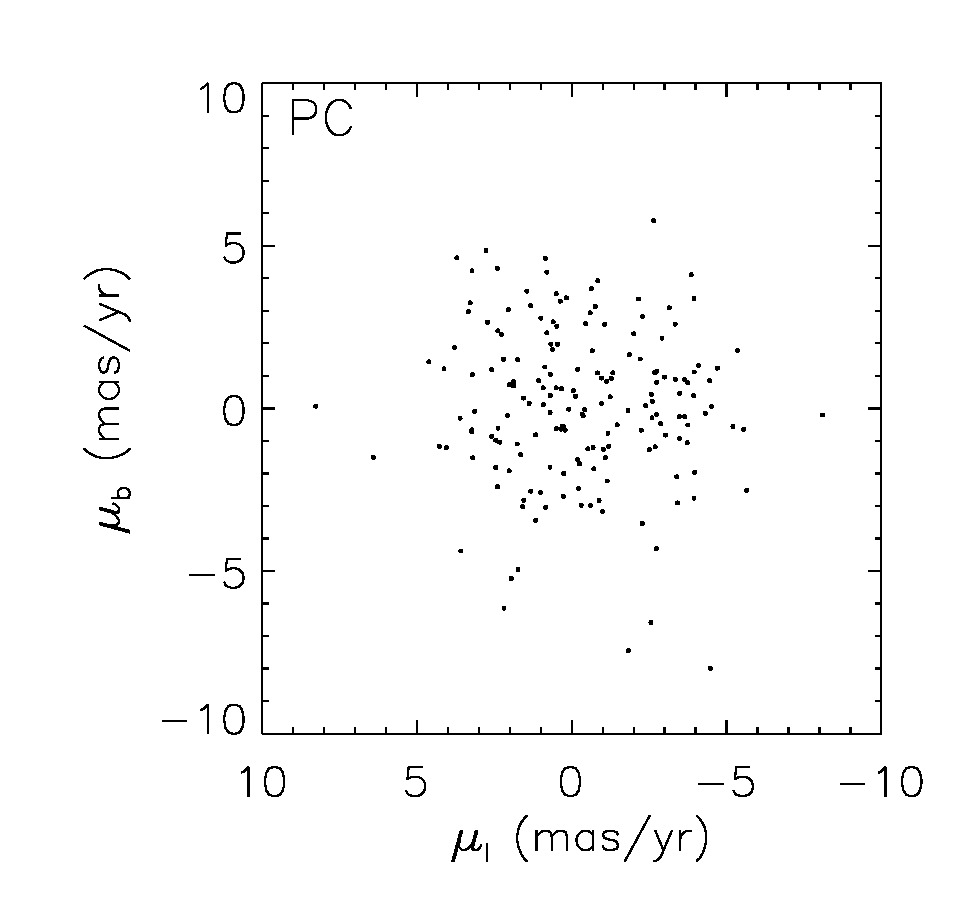}
\includegraphics[width=6.5cm, angle=0]{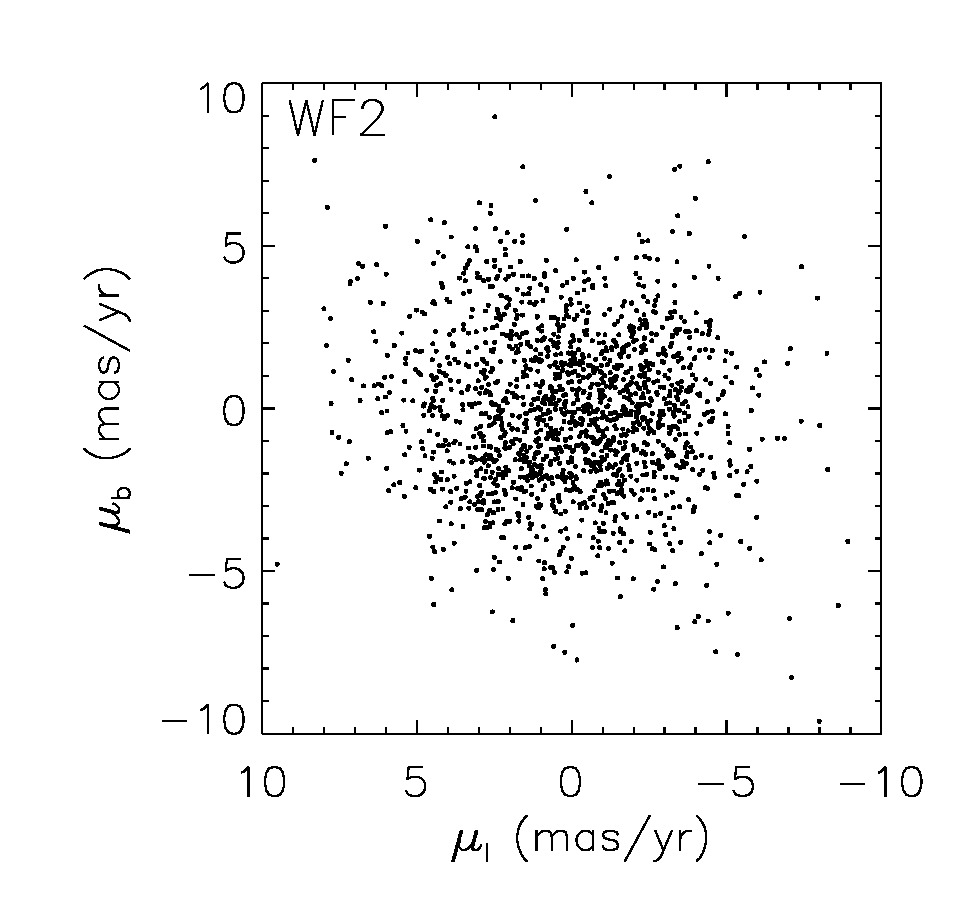}\\
\includegraphics[width=6.5cm, angle=0]{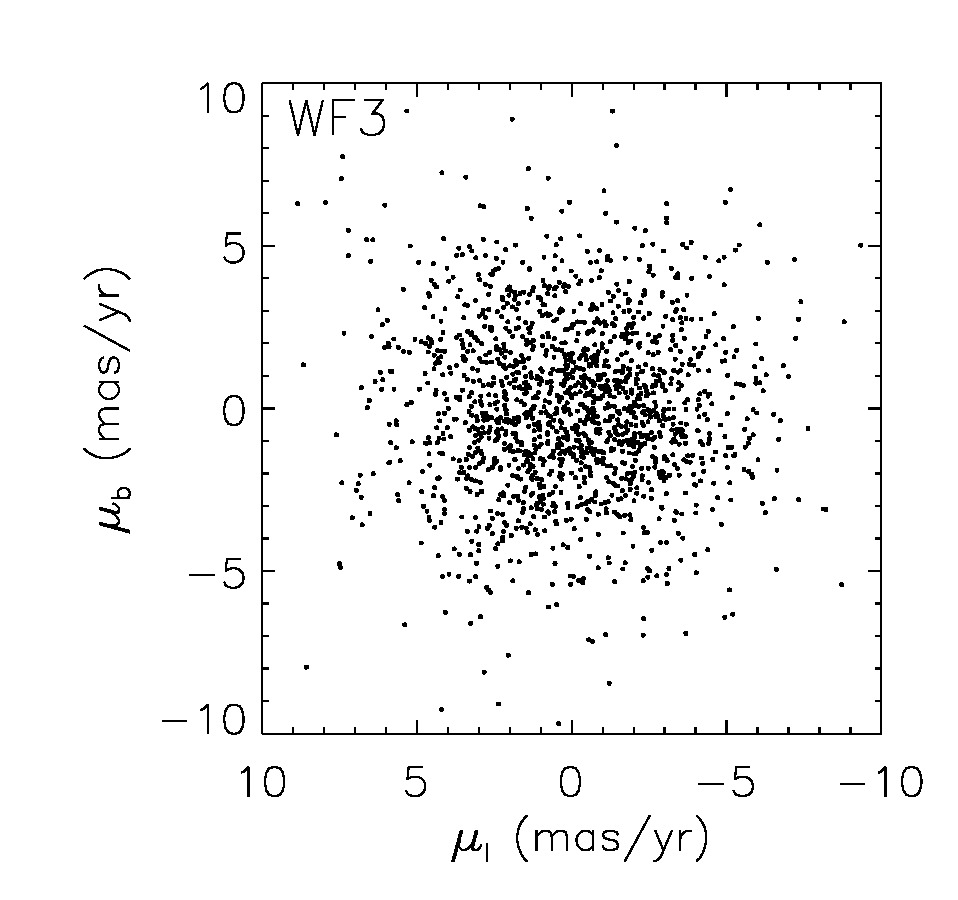}
\includegraphics[width=6.5cm, angle=0]{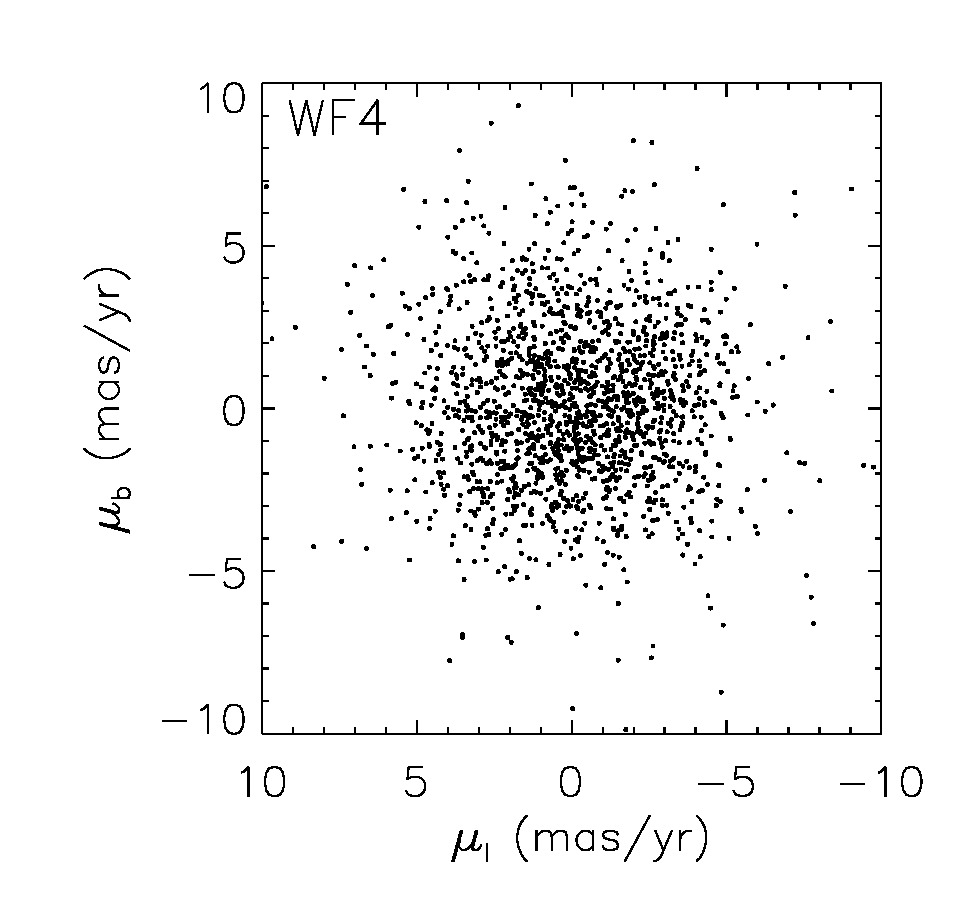}
\caption{\emph{Top row:} proper motion distribution for WFPC2 in  Field4-7.
 \emph{Second and third rows:} proper motion distribution on
  individual CCD frames of WFPC2, Planetary Camera (PC), and Wide Field
 Camera 2 (WF2), Wide Field 3 (WF3) and Wide Field 4 (WF4).
\label{fig:f182216pm}} 
\end{figure*}

\begin{figure*}
\centering
\includegraphics[width=6.5cm, angle=0]{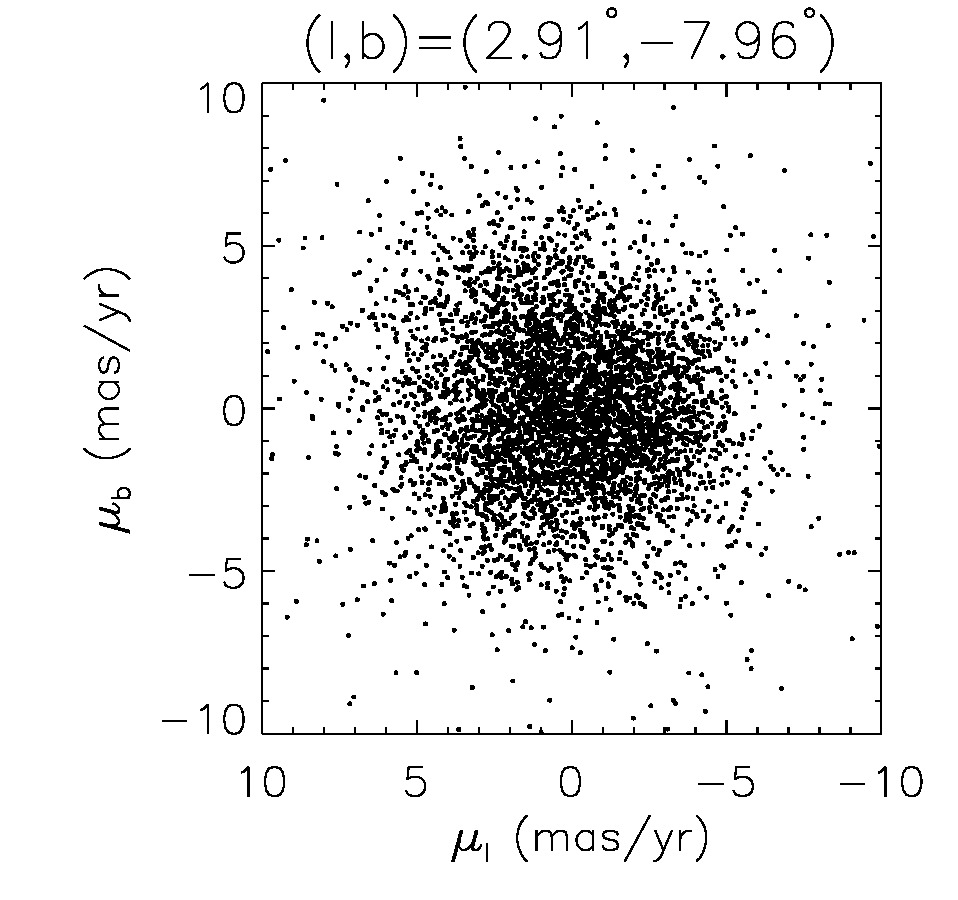}\\
\includegraphics[width=6.5cm, angle=0]{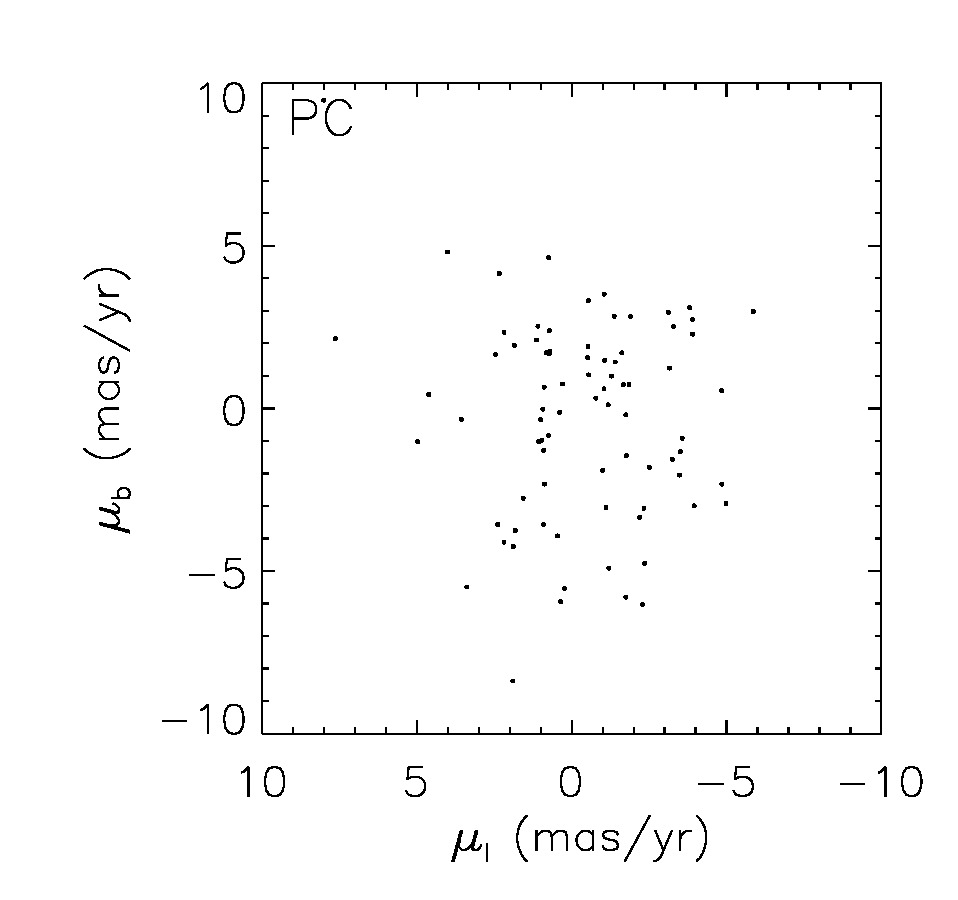}
\includegraphics[width=6.5cm, angle=0]{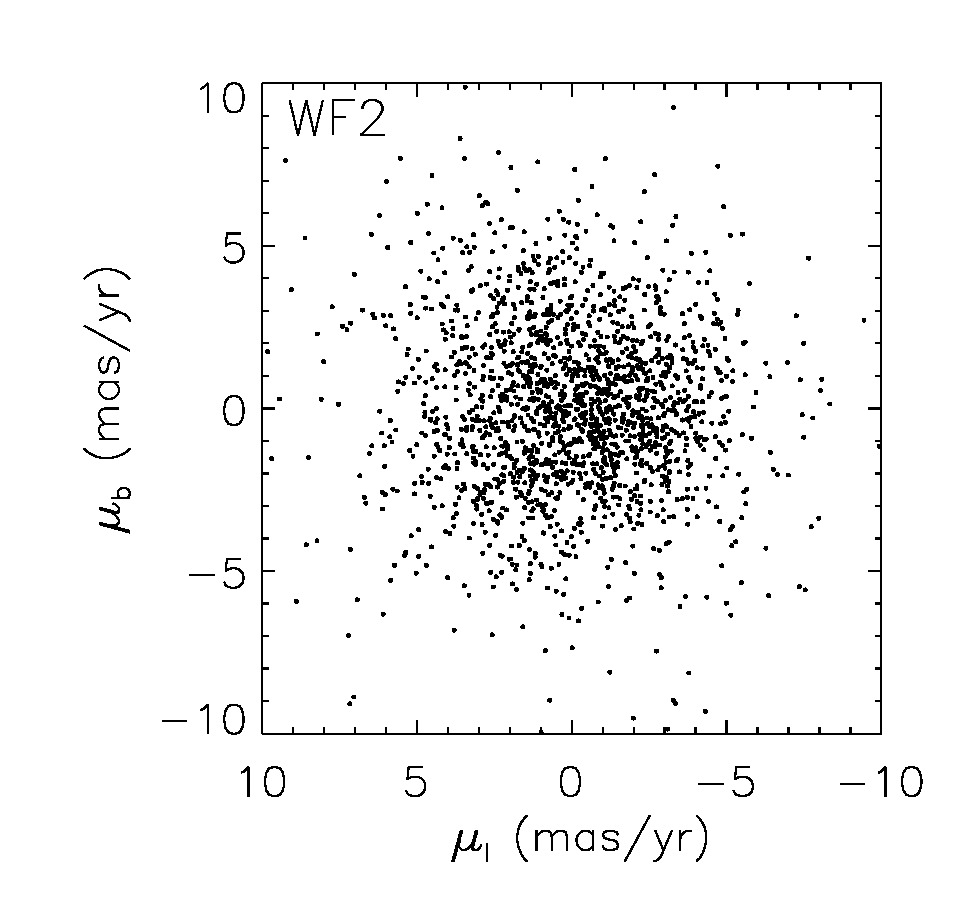}\\
\includegraphics[width=6.5cm, angle=0]{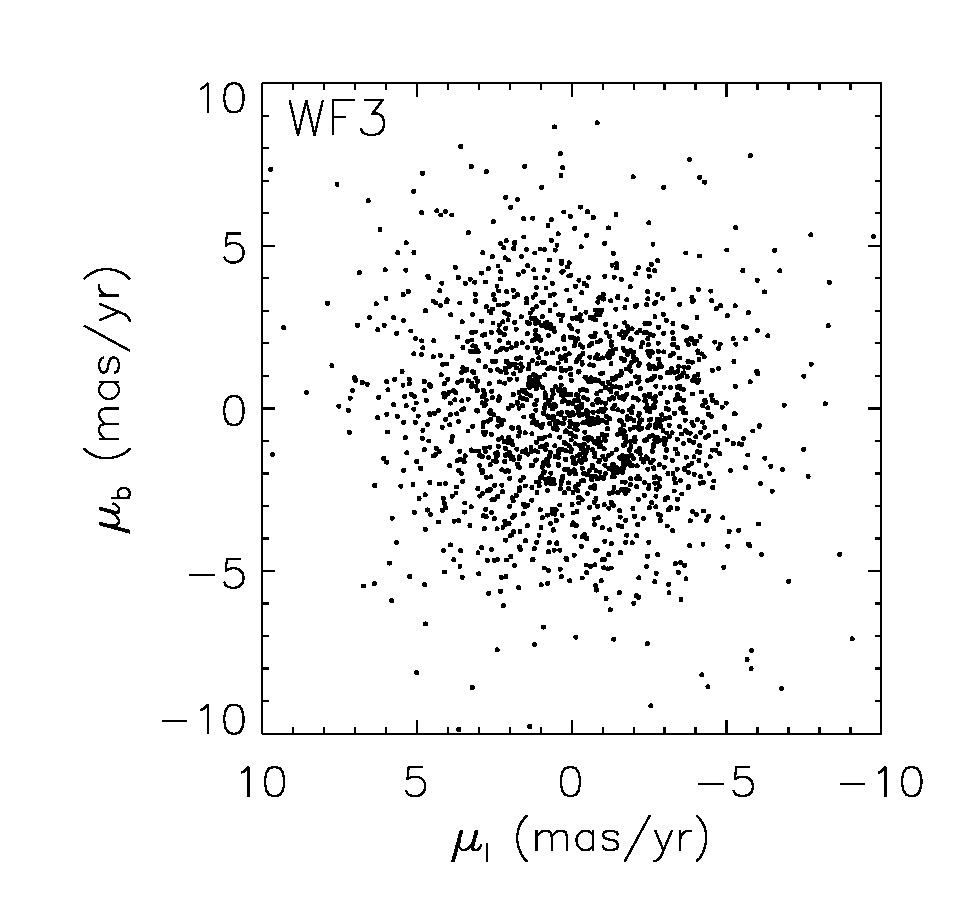}
\includegraphics[width=6.5cm, angle=0]{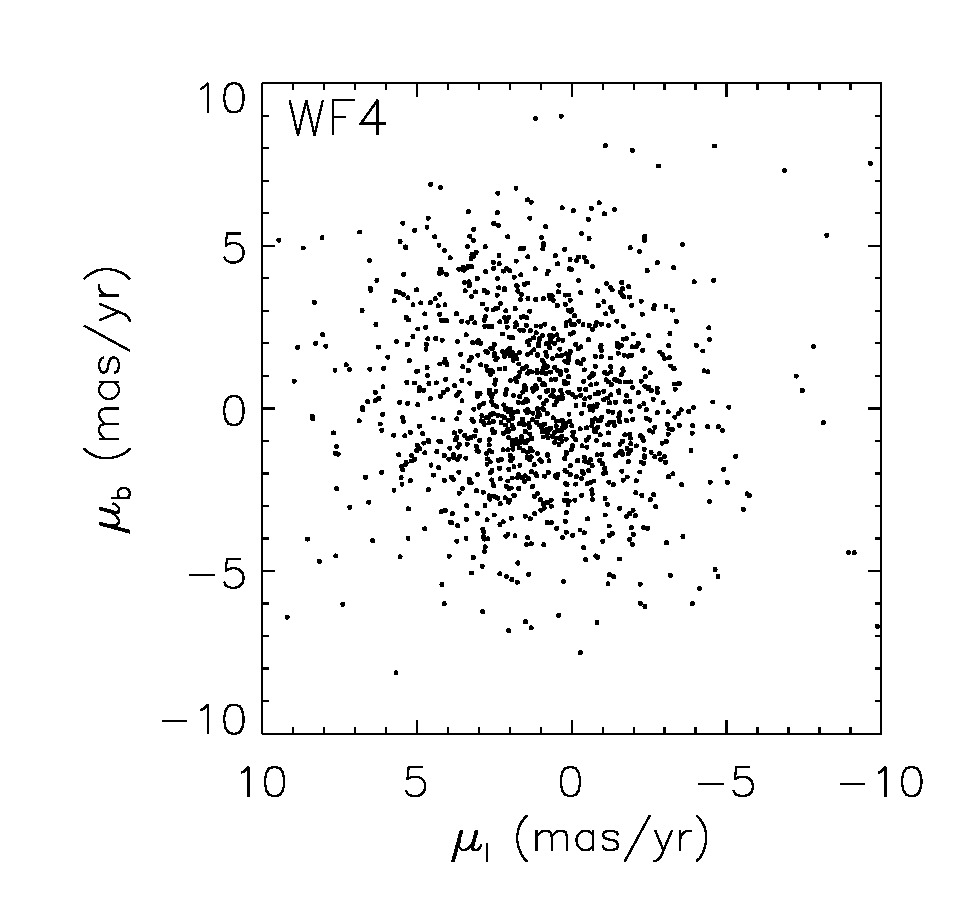}
\caption{Same as Fig. \ref{fig:f182216pm}, but for the Field 3-8   
\label{fig:f182409pm}} 
\end{figure*}

\begin{figure*}
\centering
\includegraphics[width=6.5cm, angle=0]{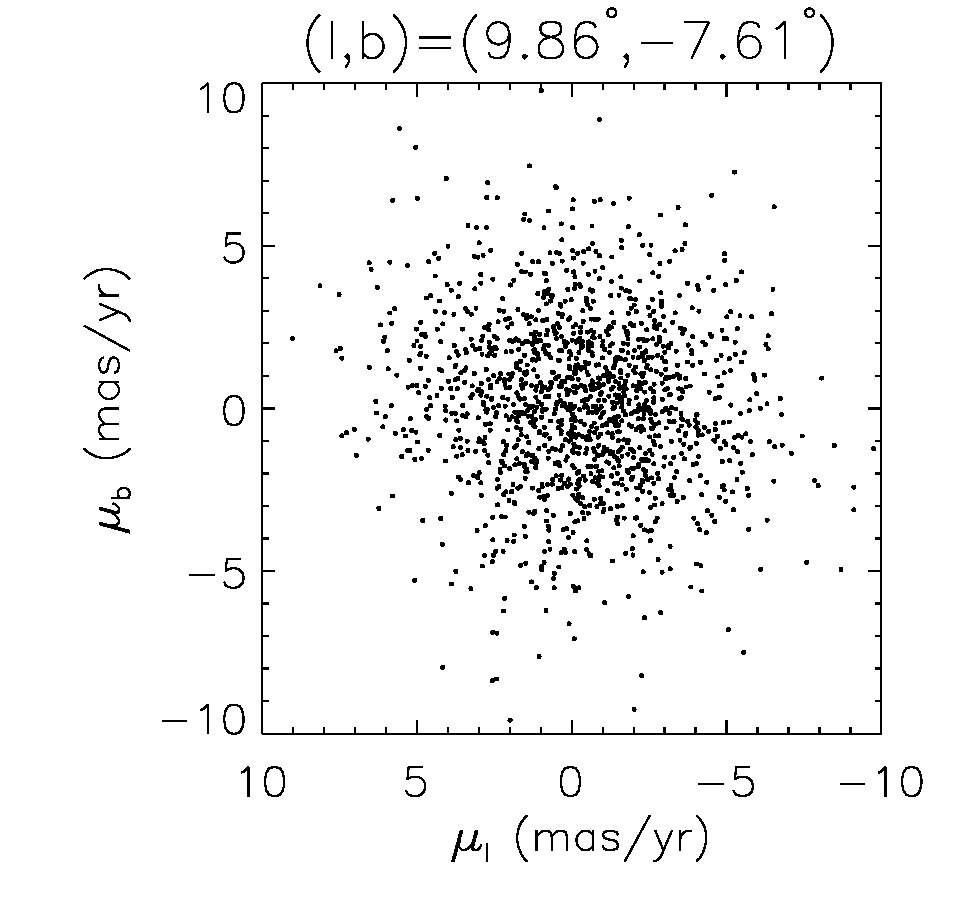}\\
\includegraphics[width=6.5cm, angle=0]{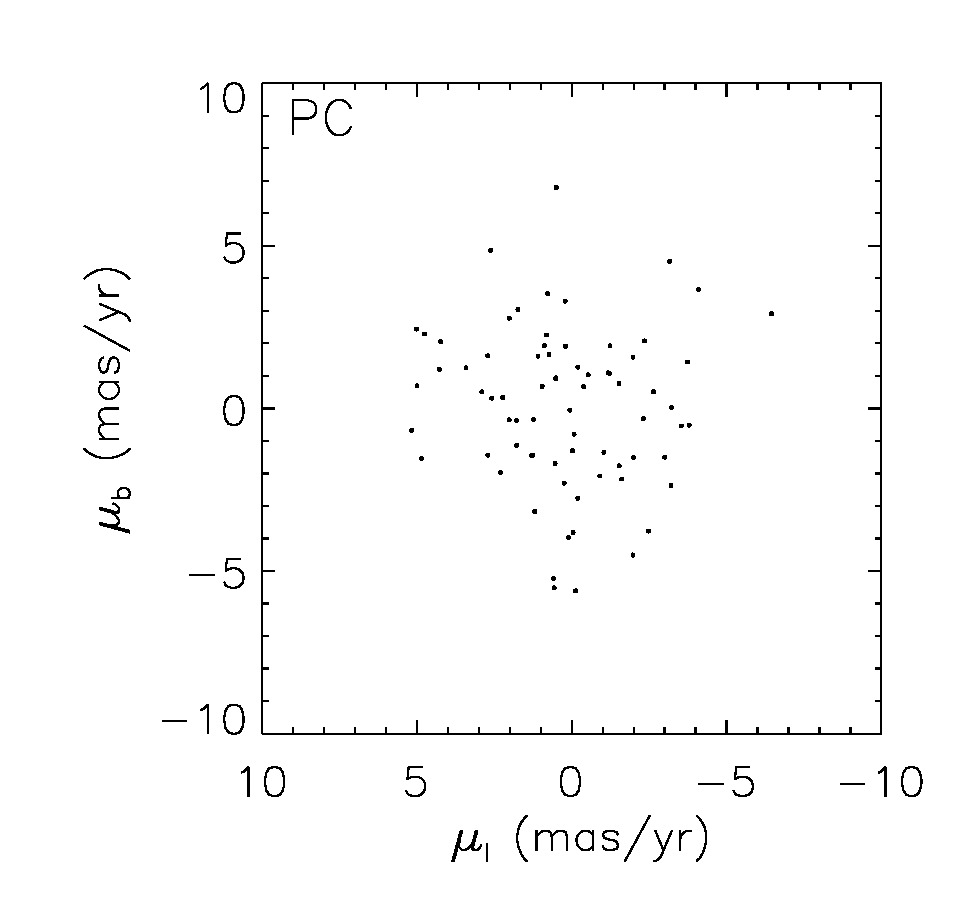}
\includegraphics[width=6.5cm, angle=0]{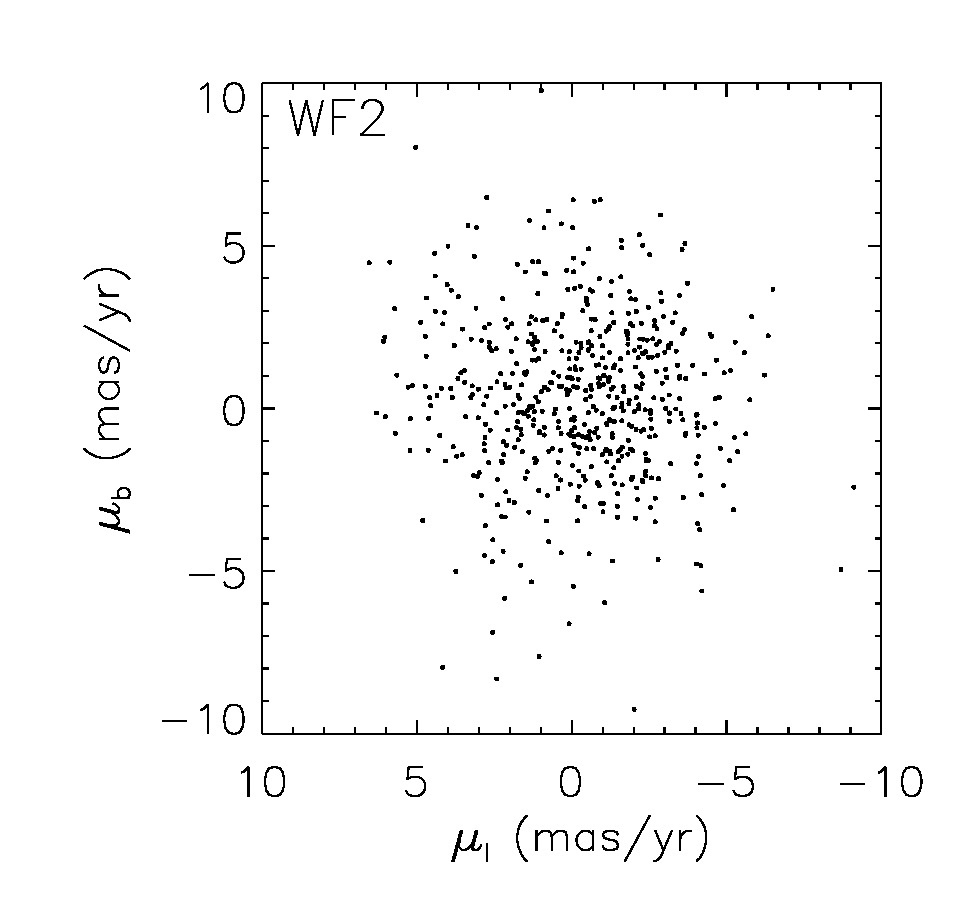}\\
\includegraphics[width=6.5cm, angle=0]{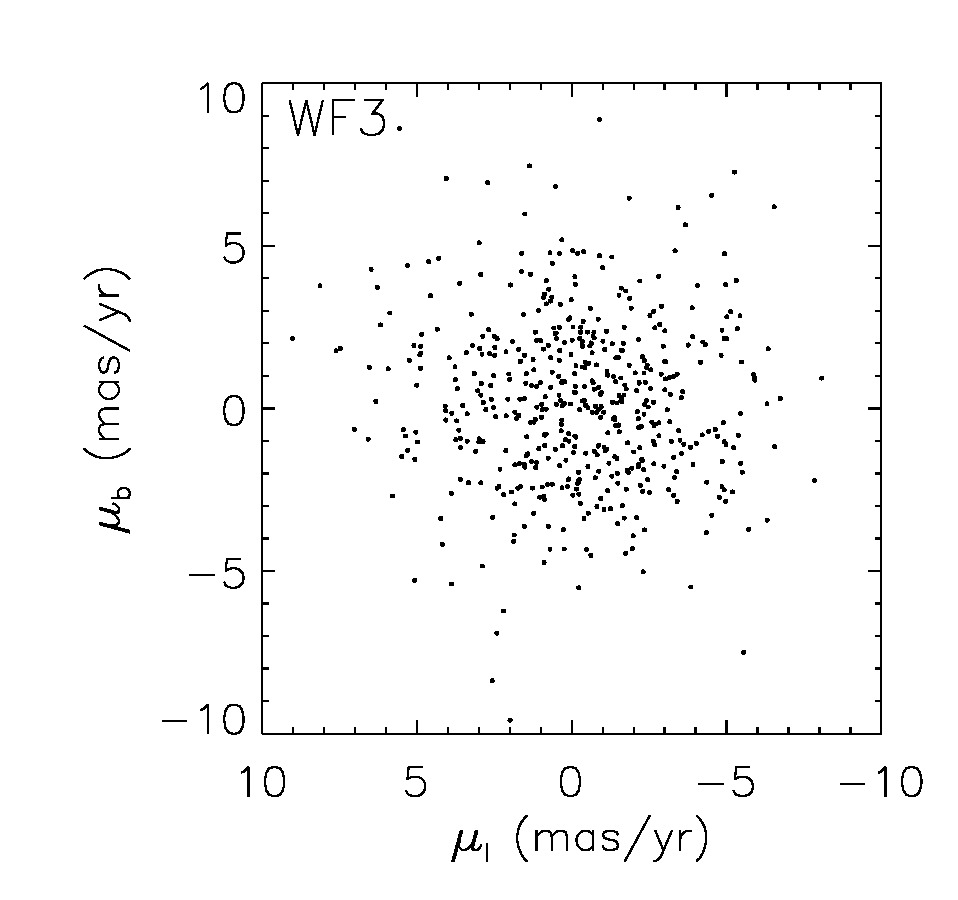}
\includegraphics[width=6.5cm, angle=0]{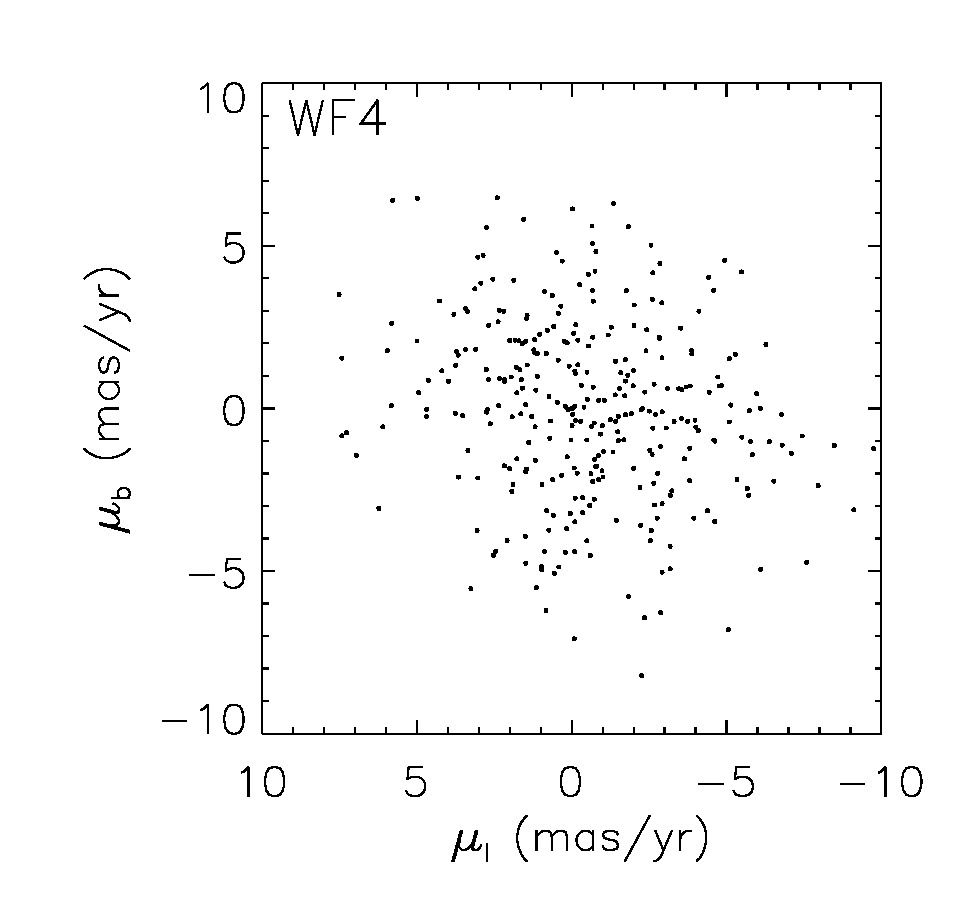}
\caption{Same as Fig. \ref{fig:f182216pm}, but for the Field 10-8  
\label{fig:f183633pm}} 
\end{figure*}

 These positions on 
 each exposure are then aligned to a common reference frame by fitting
 a polynomial to each of the sets of coordinates of  bright unsaturated stars. 
 With the positions calculated for each exposure in the
   reference frame, it is straightforward to then extract the 
 proper motions.  First the proper motions must be separated from the rest of the effects included within 
 these positions. These include a general transformation that maps the positions of 
 each exposure to those
 of the reference frame, which also includes the residuals of the geometric
   distortion of the individual frames, the proper motions, the average position residual as a function of 
 pixel phase for each image that is subtracted from each measurement,  
 and the average residual position as a function of the ``34th row'' effect 
 (\cite{anderson99}). This process is iterative, and produces a proper motion
 solution in each loop for each star. 
 This solution is obtained from a weighted linear 
 least-square fit with rejection of outliers, where the weights are estimated from the 
 centroid errors, which are in turn estimated based on photon
   noise and systematic residual errors.
 Figure \ref{fig:syserror} shows an example of the the rms residuals of the
   positions in one of our fields (Field 4-7), in the chip WF4. While
   the dominant source of error for the faint stars in the first 
   undithered epoch
  seems consistent with photon noise, the brighter stars include a
  systematic residual.  On the other hand, the second epoch (properly
  dithered) exposures show residuals consistent with a systematic error,
  which have been correctly considered in our proper motion
  solution. In both cases we have found that for the fainter end, the
  systematic errors are below $\sim 0.8\ mas\ yr^{-1}$ for a baseline
  of $9\ yr$, which at a distance
  of $8\ kpc$ corresponds to $\sim 30\ km\ s^{-1}$.  
 A more detailed account of the technique can be found
 in KR02, so will not be repeated here.

\begin{figure*}
\centering
\includegraphics[width=15cm]{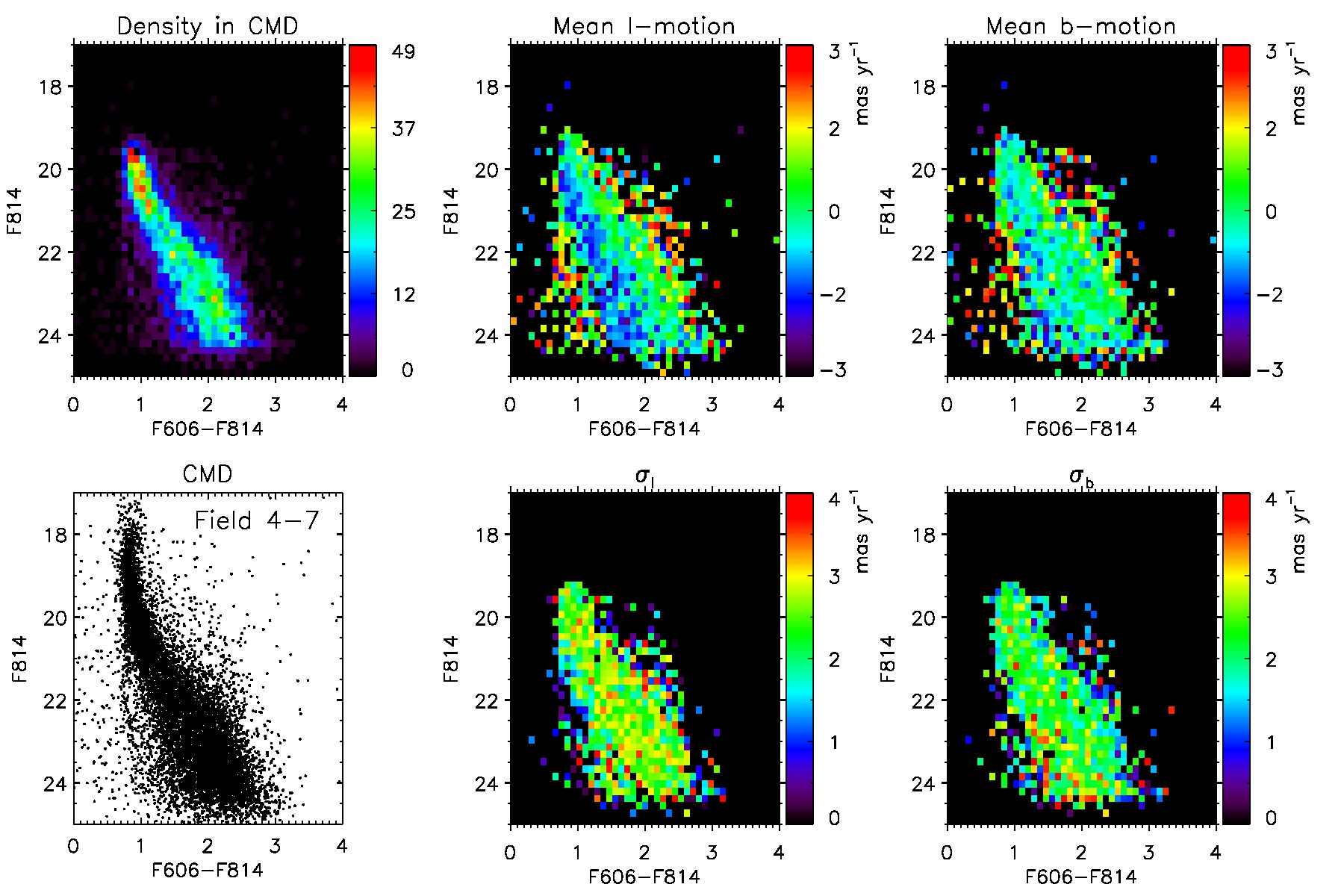}
\caption{Binned CMDs of Field 4-7 field
  $(l,b)=(3.58^{\circ},-7.17^{\circ})$ with n$_\mathrm{fit}\geq 6$.
 The unbinned CMD is the only plot in this figure including all the stars
 with available photometry regardless of the respective n$_\mathrm{fit}$ for clarity. 
\label{fig:f182216_cmd}} 
\end{figure*}
\begin{figure*}
\centering
\includegraphics[width=15cm]{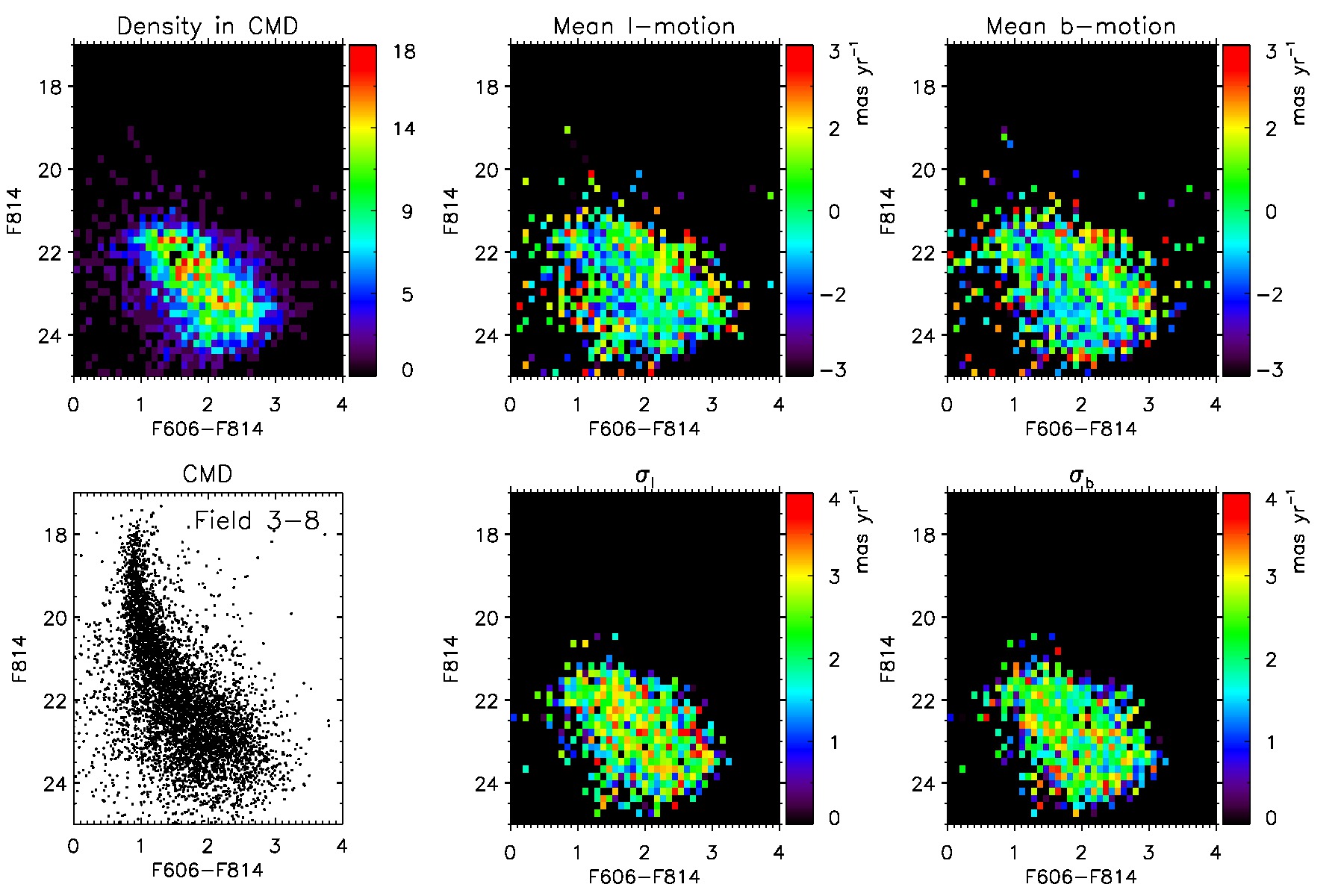}
\caption{Binned CMDs of Field 3-8 field
  $(l,b)=(2.91^{\circ},-7.96^{\circ})$ with n$_\mathrm{fit} \geq 5$  
\label{fig:f182409_cmd}} 
\end{figure*}

\begin{figure*}
\centering
\includegraphics[width=15cm]{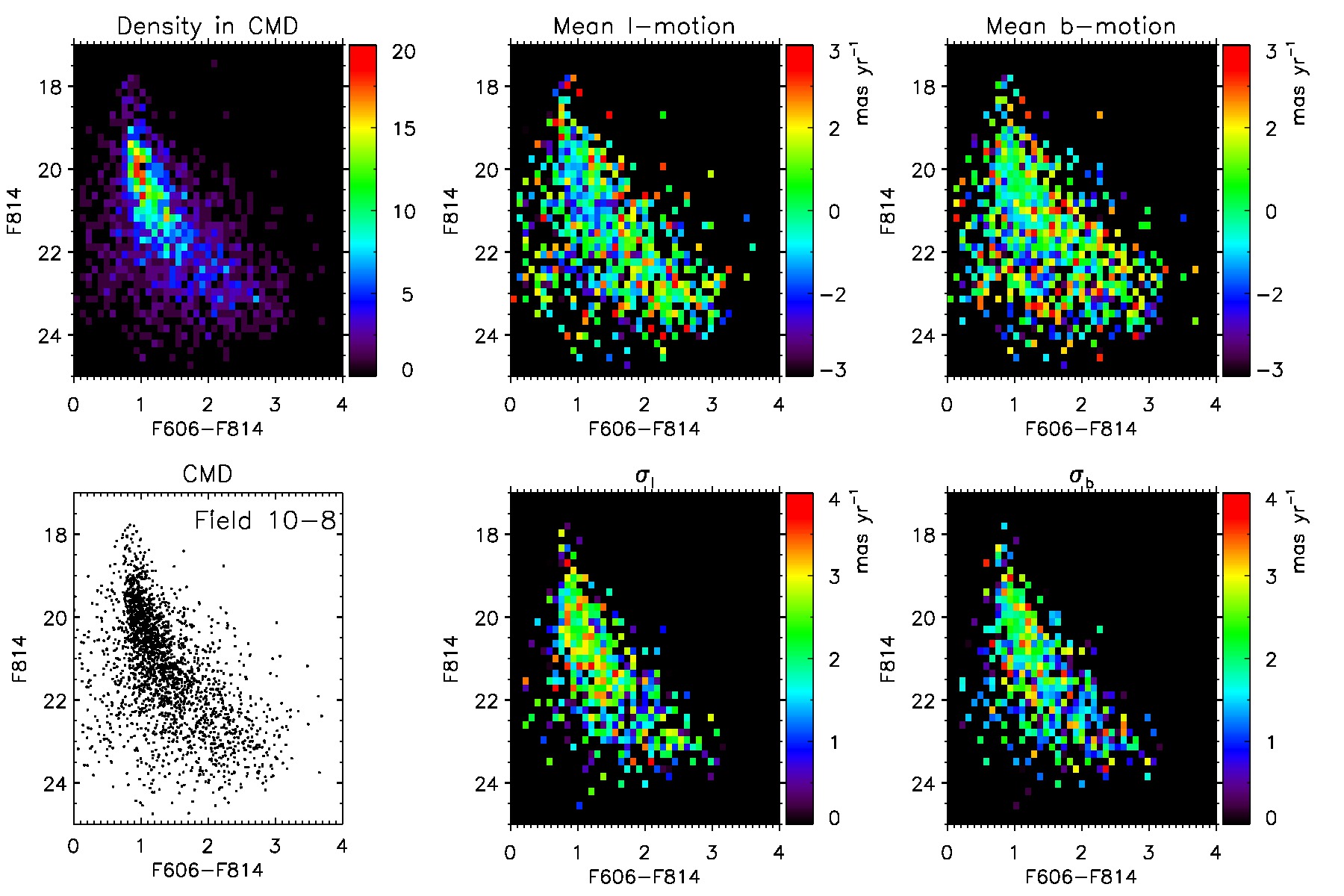}
\caption{Binned CMDs of Field 10-8 field
  $(l,b)=(9.86^{\circ},-7.61^{\circ})$ with n$_\mathrm{fit} \geq 6$     
\label{fig:f183633_cmd}} 
\end{figure*}

 We must stress that these are \emph{relative} proper motions, because
 they assume that 
 the average movement of all the stars in the field is zero. Absolute proper motions
 would require identifying of extragalactic sources to use as references in the same field. 
 The zero-mean assumption works well for bulge fields, but breaks down when 
 many stars have the same peculiar velocity.
 Such is the 
 case of the Field 10-8, where many of the bright stars in the master list correspond
 to the globular cluster NGC 6656 (M22). To make matters worse, the mean proper motion
 is different for the four chips of WFPC2 since the cluster star fraction depends on 
 position. To remove 
 this effect, the cluster stars in each image are removed from the master list by 
 identifying the cluster stars from the proper motions calculated from all 
 the stars in the field.  Taking advantage of the small proper motion dispersion of the cluster,
 the NGC 6656 stars can be easily identified (and removed). 
 The proper motions are then calculated again using only non-cluster stars, in which it is
 reasonable to use the assumption of average movement zero. 
 This procedure is illustrated in Figure \ref{fig:pm2coor_wf4} for one of the chips, WF4. 
 In a similar way, we repeated the latter procedure for 
 cluster stars in order to exclude bulge contamination.

\subsection{NGC 6656 proper motions}

 In a globular cluster, the velocity dispersion observed must be small 
 to maintain a dynamically stable structure. 
 This is consistent with the previously reported kinematics in NGC 6656;
   the \emph{Catalog of Parameters for Milky Way Globular Clusters} (Harris 1996) lists a velocity
   dispersion of $\sim$ 7.8 $km\  s^{-1}$ for NGC 6656, based on radial 
 velocities. Similarly,  Peterson \& Cudworth (1994) reported a velocity dispersion of 
 $6.6 \pm 0.8 km\ s^{-1}$ for stars within a $1'-7'$ annular field,
 and a proper motion dispersion of $0.56 \pm 0.03\ mas\ yr^{-1} $, the
 latter corresponding to $\sim 8.5 \pm 0.5\ km\ s^{-1}$ using a distance of
 $3.2\ kpc$ (Samus et al. 1995).
  More recently, Chen Ding et al. (2004) have used HST
    WFPC2 observations to obtain a proper
  motion dispersion $\overline{\sigma}=1.12\ mas\ yr^{-1}$, or
    $17\ km\ s^{-1}$. 

 As mentioned before, Field10-8 was separated into a cluster and a bulge component 
 during the proper motion 
 procedure by means of a pure kinematic selection.
 This kinematic selection was performed for each chip by selecting a
small radius around the cluster proper motions (see Figure 
\ref{fig:pm2coor_wf4}), where our WFPC2 fields sample cluster stars in a
radius between $\sim2'$ and $\sim5'$ from the center of NGC 6656. 
Table~\ref{tab:velnumdisp2} and Figure~\ref{fig:ngc6656pm} show the
 number of cluster stars selected per field and proper motion
 dispersions,
 while Figure~\ref{fig:f183633_cmd_b} shows the respective binned
   CMD. Our first impression is that these plots do not indicate significant gradients or
 variations, as expected of a globular cluster, and seem consistent with the previously listed
   literature. 
 Furthermore, the small dispersion observed in our cluster proper
  motions results, combined with
  the reported kinematics, can be used as an external 
  assessment of the accuracy  of our procedures.     
%
%
%
%
%
 Table \ref{tab:velnumdisp2} indicates a dispersion in $l$ and $b$ of 
 $\sim 1.04\ mas\ yr^{-1}$  for cluster stars at the WF
 chips, which corresponds to $\sim 17\ km/sec$. The PC dispersions are
 about twice as small, a fact that we interpret as the direct effect
 of our undithered first-epoch pixel size, as the source of systematic
 errors in the measurements. 
  By substracting in quadratures the Peterson \& Cudworth (1994) determination
  of the intrinsic velocity dispersion to our values in
 Table \ref{tab:velnumdisp2}, we obtain a velocity error of   $\sim 0.87\ mas\ yr^{-1}$ for
 the WF chips. Repeating the same analysis in  our NGC 6656 PC proper motions
 yields precisions consistent with $\sim 0$ error, which suggests an
 overestimation of the previously reported intrinsic proper motion dispersion in NGC 6656, 
 and requires further analysis.

 Thus, our results confirm the reliability  of our proper motion technique and in addition,
 match the precision achieved by our bulge radial velocities
 measurements in our three off-axis
 fields ($\sim 30\ km\ s^{-1}$), which are included in a separate paper
 (Soto et al. 2012).

\section{Analysis}
 Our proper motion results are plotted in Figures~\ref{fig:f182216pm}, 
 \ref{fig:f182409pm}, and \ref{fig:f183633pm}; the proper motion 
 dispersions, and numbers
 per field are given in Table~\ref{tab:velnumdisp}, where the
   terms of the velocity ellipsoid tensor (Zhao et al. 1994) are defined by
\begin{equation}
\sigma_{ij}^2=\frac{N}{N-1} (\langle V_i V_j \rangle - \langle V_i \rangle \langle V_j \rangle ). 
\end{equation}
 Color-magnitude diagrams for the three
 fields color-coded by the proper motion information are shown in Figures 
 \ref{fig:f182216_cmd}, \ref{fig:f182409_cmd}, and \ref{fig:f183633_cmd}. 
 The first indication of the correct performance of our procedures appears in 
 Table \ref{tab:velnumdisp} in that there is no evidence of significant variations or
 inconsistencies in the kinematics between the stars in the WF and PC chips.
 Similarly, we find that 
 the proper motion distributions do not differ dramatically from one field to the other.  This agreement
 in the proper-motion distribution seems to even extend to the fields presented in 
 KR02 close to the Galactic minor axis (see Table \ref{tab:konraddisp}).

\begin{table}
\begin{center}
\caption{Proper motion dispersions in Kuijken \& Rich (2002)}
\label{tab:konraddisp}
\begin{tabular}{l c c c c c}
\hline \hline
Field    & N & l & b & $\sigma_l$ & $\sigma_b$ \\
         &   & & & $(mas \ yr^{-1})$ & $(mas \ yr^{-1})$ \\
\hline
BW  & 15862 &$1.^{\circ}13$& $-3.^{\circ}77$& 2.98$\pm$0.02 & 2.54$\pm$0.01 \\
Sgr-I       & 20234 &$1.^{\circ}25$& $-2.^{\circ}65$ & 3.24$\pm$0.02 & 2.77$\pm$0.01 \\
\hline
\end{tabular}
\end{center}
\end{table}
\normalsize

 The three fields in the present study, all of which are at positive Galactic longitudes,
 have distributions with
 no significant correlation $r_{lb}$ between $l$ and $b$ proper motions; at the same time, $\sigma_l$    
 and $\sigma_b$ are slightly larger than the dispersions found in minor axis fields 
 in KR02. 
 Zoccali et al. (2008) found a decreasing dispersion of radial
  velocities when moving away from the disk for a bulge population close
  to the minor axis. A similar result was found by Soto et al. (2012) in their
  minor axis fields, consistently. This gradient is not observed in the radial
  velocities of the same off-axis fields analyzed in this work.  
 There are two possible reasons for an increase in $\sigma_l$ and $\sigma_b$.  
 It either may be that the increase in dispersion is due to 
 a higher contamination by foreground disk populations, thereby decreasing 
 the total number of bulge stars detected or that there is a real, intrinsic increase 
 in the dispersions due to the location of the fields in the sky
 (e.g., compared with the minor axis fields, we might be
   sampling different bulge orbit families or the same orbits
 at a different bulge location in this off-axis fields). 
 Below we try to discern which of these effects prevail using the assistance of the 
 CMD information.

\begin{figure}[!h]
\centering
\includegraphics[width=6cm, angle=0]{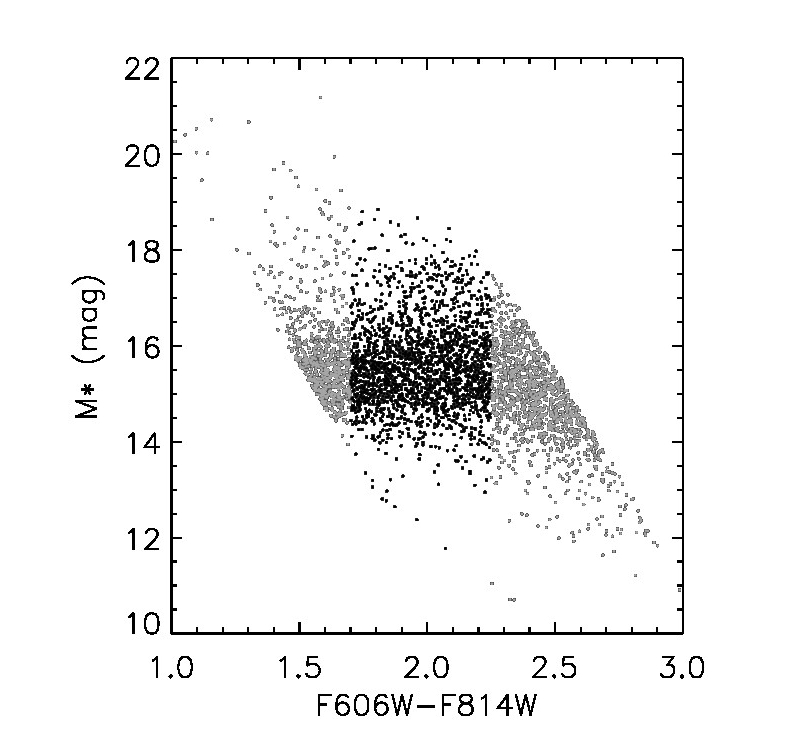}
\caption{Photometric parallax M* as a function of the color
  for a subsample of main-sequence stars  in Field 4-7. Black
  dots have been used to derive the respective photometric
  distance D*.
\label{fig:f182216_dbin_p1}} 
\end{figure}

 The binned CMDs (Figures~\ref{fig:f182216_cmd}, \ref{fig:f182409_cmd}, and \ref{fig:f183633_cmd}) 
 for the three fields show some distinctive features. 
 First, there is a cut-off at the bright end of the CMDs because
 of saturation in the first-epoch long exposures ( $> 1200 s$).
 This saturation limit, 
 in turn, limited our proper motions to main sequence (MS) stars below 
 the turn-off.  In contrast, the fields in KR02 had many shorter exposures in their first epoch,
 allowing them to reach both the turn-off and red-giant branch (RGB). A second  
 feature hinted in the three fields is 
 a gradient perpendicular to the main sequence in the mean $\mu_l$. 
 Such a gradient, 
 which for a given color shows a drift toward negative $\mu_l$, has
 been clearly observed in
 the Galactic minor-axis fields in KR02.  In our fields, however, the appearance is noisier, which is likely
 due to the reduced statistics. The gradient can be explained as a signature of the 
 rotation through the galaxy, 
 where high positive $\mu_l$ should correspond mainly
 to the foreground population rotating in front of the bulge. 
 We explore 
   the significance of this gradient later in this section in more detail. 
Interestingly, a similar
 but noisier feature appears in the mean $\mu_b$ panels.  This might be caused 
 by a combination of the projection effect of the bulge orbits and the
 contamination by the disk.

\begin{figure*}
\centering
\includegraphics[width=12.0cm, angle=0]{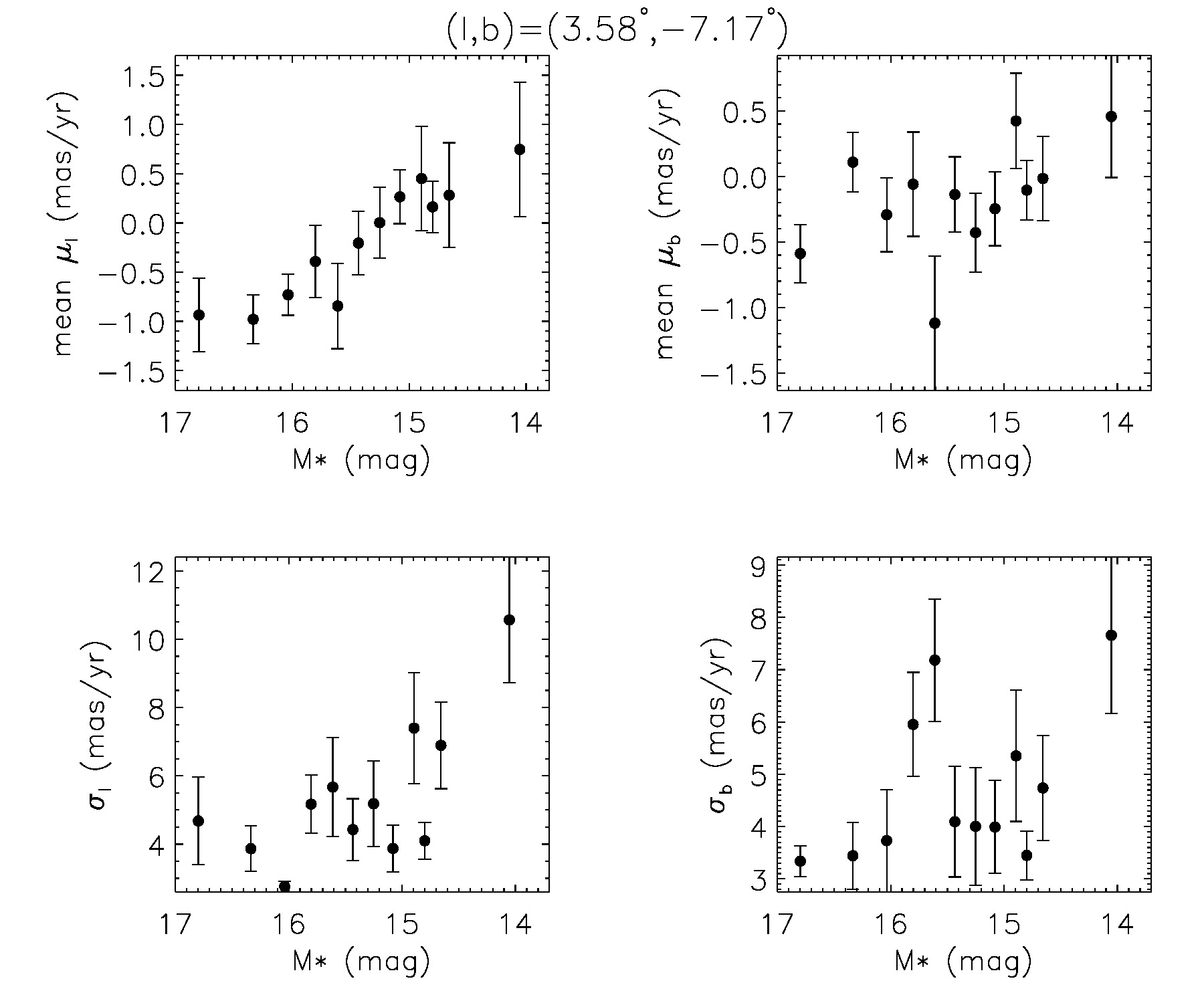}
\caption{Mean proper motions and dispersions as a function of the photometric parallax $M*$,
 for the subsample of main sequence stars in Field 4-7, where each
 point corresponds to 180 stars. The higher number of stars on this
 field has allowed us to explore  the proper motions as a function of 
 M*. The rotation pattern speed clearly appears on this field for $\mu_l$
 (\emph{top left}).  Similarly, $\sigma_l$ (\emph{bottom, left}) also
 indicates a decreasing dispersion for stars lying farther away, as
 expected when observing the Galactic rotation. 
\label{fig:f182216_mumstar}} 
\end{figure*}

 We explore these effects further on Field 4-7, which consists of
 a higher number of proper motions compared with Field 3-8 and
 Field 10-8, and therefore allows a more 
 robust analysis. 
 We followed a similar procedure to dissect the stellar populations in our fields
 than KR02 and K04.
  We have chosen a crude distance modulus as an indicator of the
  distances toward the Galactic bulge. The quantity M* can be defined as
\begin{equation}
 M*= m_{814W} - 3 \times (m_{814W} - m_{606W}),
\end{equation} 
 and has been used to remove the slope of the main sequence stars in the CMD for
 our proper motion sample. 
The coefficient of three for the color, as opposed to two in KR02, is due to
the use of F606W-F814W, as opposed to F555W-F814W.
 Thus, the proper motions as a function
 of this distance modulus, M*, can be explored. 
 Figure \ref{fig:f182216_dbin_p1} shows the distance modulus
   M* as a function of the color for a subsample of main-sequence stars, while
 Figure
 \ref{fig:f182216_mumstar} shows the proper motion means and dispersions, as a function of M* for Field 4-7. 
 Again, this field was chosen for its better statistics as compared
 to the number of stars with proper motions in the two other fields, Field 3-8 
 and Field 10-8. 
 The mean $\mu_l$ motion again shows the kinematic feature
   previously observed in Figure \ref{fig:f182216_cmd}, which we
   relate to the rotation of stars across the Galactic bulge. In
   addition,   
   Field 4-7 proper motion dispersions $\sigma_l$
 show a mild trend for closer main sequence stars to have 
 comparatively higher values than main sequence stars lying
 farther away. This can be interpreted as a distance effect, and 
 it is also a nice demonstration that noise does not dominate our
 results.

\begin{figure*}
\centering
\includegraphics[width=10.6cm, angle=0]{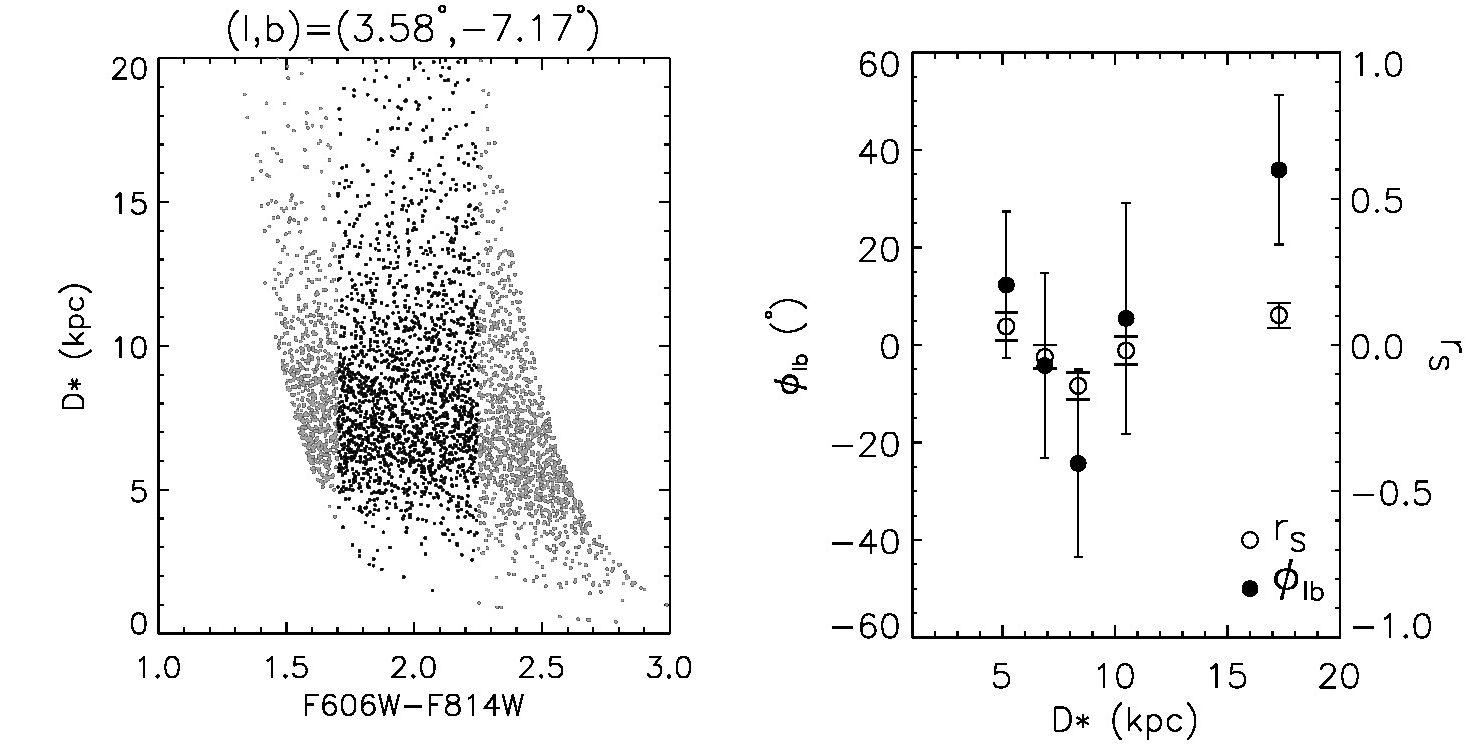}
\caption{
\emph{Left,} Distance D*, derived from M*, as a function of the
color in Field 4-7. \emph{Right,} Angle $\phi_{lb}$ 
 as a function of binned D*, equal numbers of stars have been selected per bin; in addition, the
 Spearman correlation coefficient $r_S$ has been calculated in each bin to have a independent correlation measurement between $l$ and $b$. 
\label{fig:f182216_dbin_a}} 
\end{figure*}

\begin{figure*}
\centering
\includegraphics[width=15.0cm, angle=0]{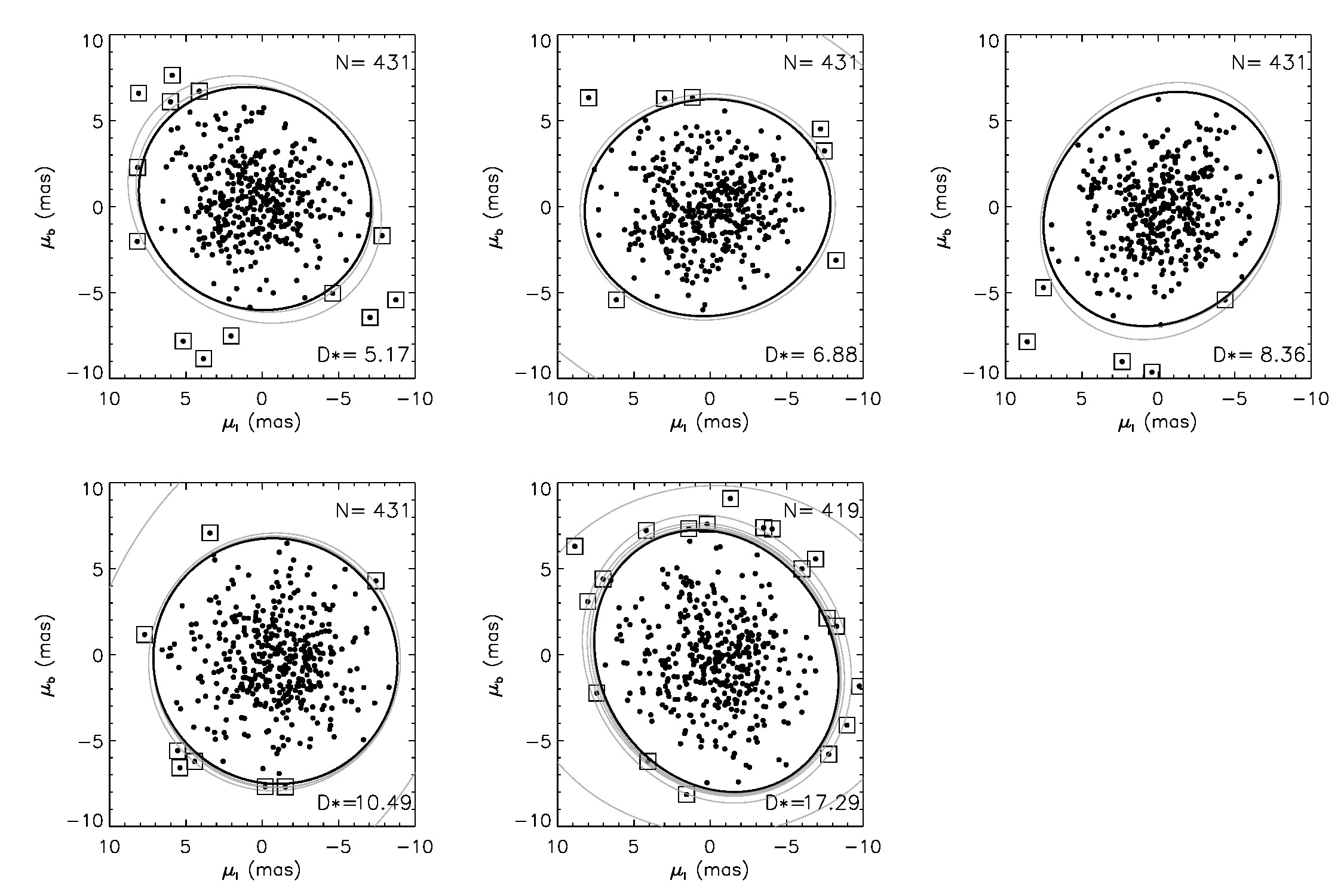}
\caption{Velocity ellipsoids used to calculate the angle $\phi_{lb}$ in Field 4-7. An iterative 
clipping algorithm has been used to exclude stars beyond $3\sigma$. Rejected stars during the process are enclosed by squares.   
\label{fig:f182216_dbin_b}} 
\end{figure*}

 Figure~ \ref{fig:f182216_dbin_a} illustrates the 
photometric distance D* (left panel), derived from M*, as a function of the color.     
 The right panel shows the 
 angle $\phi_{lb}$ as a function of distance, D*.  Here , $\phi_{lb}$ is the equivalent to the 
 vertex angle, but in the $l$-$b$ velocity plane, and D* has been binned for clarity. 
 In order to obtain completeness in distance reaching our last bin (D*$\sim$20 kpc), 
 a narrower region in D* was selected from the initial main sequence subsample.
 The selection is based on the one performed in K04 for the field \emph{near} NGC 6558, 
 and is shown by the black points in the Figure
 \ref{fig:f182216_dbin_p1} and the left panel of the 
 Figure \ref{fig:f182216_dbin_a}.  The angle $\phi_{lb}$ has
 been calculated using an iterative clipping algorithm that rejects stars beyond
 $3\sigma$ of the distribution to avoid contamination. In addition, 
 the Spearman correlation coefficient, $r_S$, is calculated in each
 case from the same stars in the first iteration of the
   velocity ellipsoid calculation, in order to have 
 an independent measurement of the correlation between $\mu_l$ and $\mu_b$.  
 Errors in $\phi_{lb}$ and $r_S$ come
 from bootstrap Montecarlo realizations, and the error bars are indicated in the Figure~\ref{fig:f182216_dbin_b}.

\begin{table}
\begin{minipage}{\linewidth}
\renewcommand{\footnoterule}{}
\caption{$\phi_{lb}$ and Spearman correlation coefficients $r_S$ at
  different distance bins for Field 4-7}
\label{table:rscoef}      
\centering                          
\begin{tabular}{l l c r c}        
\hline
\hline
D* & N & $\phi_{lb}$ &$r_S$  \footnote{Spearman's correlation coefficient} \ \ \ & $Prob(r_S)$ \footnote{Significance of the correlation} \\    
(kpc) & & ($^{\circ}$) & & \\
\hline                        
5.17  & 431  &  12  $\pm$ 13   & 0.066   &   0.169 \\
6.88  & 431  &  -4  $\pm$ 12   & -0.043  &    0.210 \\
8.36  & 431  & -24 $\pm$  19  & -0.146  & 1e-7   \\
10.49 & 431 &   5   $\pm$ 9    & -0.020   &   0.407 \\
17.29 & 419 &  35  $\pm$ 14   & 0.107  &  7e-7  \\
\hline                                   
\end{tabular}
\end{minipage}
\end{table}

 Figure~\ref{fig:f182216_dbin_b}  and Table~\ref{table:rscoef} show  
 the velocity ellipsoids and respective $\phi_{lb}$ and 
  $r_S$ for Field 4-7. These parameters clearly show a 
 change in the orientation of the velocity ellipsoid. 
 Clarkson et al. (2008) also find significant changes in the 
 orientation of the ${l,b}$ velocity ellipsoid in their
 improved sample of ACS proper motions in
 Sagittarius-I.
 This orientation changes in the velocity ellipsoid are
  difficult to interpret, especially at different longitudes, owing to
  projection effects, and will be addressed in the future using our 
 dynamical models.

\section{Conclusions}
 We have presented  \textbf{$\sim 15000$}  new proper motions for three off-axis fields of the Galactic 
 bulge. The results for these three fields show remarkable agreement with the results in KR02, and thus suggests a bulge structure where the kinematics observed close to
 the center along the Galactic minor axis are repeated to some extent in higher longitudes. 
 Despite the reduced number of proper motions in comparison with minor axis fields, the 
 rotation of the bulge is still visible in our fields, which reach $l\sim 10^{\circ}$.  
 We explored the possible
 changes in the velocity (proper motion) ellipsoid within Field 4-7 as
 a function of the distance along the line of sight; as is the case 
 with the results of Clarkson et al. (2008), we found a change in the tilt of the \{l,b\}
 velocity ellipsoid. 
 
 All of this suggests that a significant fraction of the population
 follows bulge-like orbits, even at the location of our three off-axis
 fields in the outer bulge. 
 If we consider the anisotropies produced
 by the bar in the minor-axis (inner bulge) fields, what we
 observe should therefore be part of the 
 Galactic bar. The importance of the extent of the bar can be related to the evolutionary stage 
 of the bulge, where slow-rotating long bars can evolve from
   rapid-rotating  short bars secularly (e.g. Combes 2007; Athanassoula 2005), thus this
 information  can provide important constraints on the bulge
   structure and formation. 
 Dynamical models including this
 new proper motion data will be able to provide new insight into the actual bulge 
 structure, until now poorly constrained.  
   
 Finally, we have demonstrated the technical feasibility of proper-motion
 measurements using different  
 cameras with different geometries for the first and second epochs. 
Field 10-8 with its 
 globular cluster NGC 6656 
 has provided us with a direct assessment of the accuracy of our
 proper motion procedure.  
 A cluster dispersion of $\sim 0.9 \ mas\ yr^{-1}$ or $\sim \ 14 \ km/sec$ at 3.2 $kpc$
 supports our claim of $\sim30\ km\ s^{-1}$ accuracy for bulge stars.
 Unfortunately, our results are severely affected by the saturation of
 the long first-epoch exposures, and could be significantly improved by a
 third ACS epoch.
 

\begin{acknowledgements}
 MS acknowledges support by Fondecyt project 3110188 and Comit\'e
 Mixto ESO- Chile. 
 RMR acknowledges funding from GO-9816 and from
 AST-0709479 from the National Science Foundation.
 This material is based upon work supported in part
 by the National Science Foundation under Grant No. 1066293 and by the
 hospitality of the Aspen Center for Physics.
\end{acknowledgements}

\end{document}